\documentclass[two, aps,prd,twocolumn,superscriptaddress,showpacs,preprintnumbers,amsmath,amssymb]{revtex4-1}
\usepackage{graphicx,color}
\usepackage{dcolumn}
\usepackage{colordvi} 
\usepackage{bm}
\usepackage{txfonts}
\usepackage{amsmath,amssymb,amsfonts}
\usepackage[]{color}

\begin{document}
\title{Scalar Stochastic Gravitational-Wave Background in Brans-Dicke Theory of Gravity}
\author{Song Ming Du}
\email{smdu@caltech.edu}
\affiliation{TAPIR, Walter Burke Institute for Theoretical Physics, California Institute of Technology, Pasadena, California 91125, USA}


\begin{abstract}

We study the scalar stochastic gravitational-wave background (SGWB) from astrophysical sources, including compact binary mergers and stellar collapses, in the Bras-Dicke theory of gravity.  By contrast to tensor waves, we found the scalar SGWB to predominantly arise from stellar collapses.  These collapses not only take place at higher astrophysical rates, but emit more energy.  This is because, unlike tensor radiation, which mainly starts from quadrupole order, the scalar perturbation can be excited by changes in the monopole moment.  In particular, in the case of stellar collapse into a neutron star or a black hole, the monopole radiation, at frequencies below 100\,Hz, is dominated by the memory effect. At low frequencies, the scalar SGWB spectrum follows a power law of $\Omega_\text{S}\propto f^\alpha$, with $\alpha = 1$. We predict that $\Omega_\text{S}$ is inversely proportional to the square of $\omega_{\rm BD}+2$, with $\quad(\omega_{\rm BD}+2)^2\Omega_S(f=25\,{\rm Hz}) = 2.8\times 10^{-6}$. We  also estimate the detectability of the scalar SGWB for current and third-generation detector networks, and the bound on $\omega_{\rm BD}$ that can be imposed from these observations.  
\end{abstract}

\maketitle

\section{Introduction}

The first direct detection of gravitational waves (GWs) from the merger of binary black holes (BBHs) by the LIGO-Virgo collaboration~\cite{gw150914} marks the beginning of gravitational-wave astronomy and opens up a new window to the Universe.  Since then, more GW events, both from BBH mergers and from binary neutron star (BNS) mergers, are detected by the Advanced LIGO/Virgo network \cite{gw151226, gw170104, gw170814, gw170817}.  Besides these resolvable, individual GW sources we have discovered so far, a Stochastic Gravitational-Wave Background (SGWB), which arises from the population of unresolved GW events at larger distances, is anticipated to be detectable in the upcoming years \cite{gw150914implications, gw170817implications}. 

Gravitational wave signals provide us with unprecedented opportunities to test general relativity (GR) and study modified theories of gravity \cite{gw150914, gw170104, gw150914test, gw170817grb170817a}. One significant prediction of general relativity is that gravitational waves only contain two tensor polarization modes ($+$ mode and $\times$ mode). 
On the other hand, additional polarization modes  do exist in modified theories of gravity; if directly detected, they would become a strong evidence for extensions to Einstein's original theory \cite{eardley1, eardley2}. 
For example, the Brans-Dicke (BD) theory \cite{brans}, which minimally extends Einstein's gravity by incorporating a scalar field (Brans-Dicke field) coupled to the metric tensor, predicts the existence of a transverse scalar polarization mode (also referred to as the breathing, or the  ``$\circ$ mode''). Previously, several detection strategies for the non-tensorial SGWB have been proposed~\cite{maggiore, nishizawa, callister}. A recent study \cite{asearchfor}, based on the method in \cite{callister} and the data from LIGO's O1 observing run, has placed the first constraints on the contributions from non-tensorial polarizations to the SGWB. 

All the works so far have assumed general, power-law models for the energy spectra of the non-tensorial SGWB --- without considering its specific origins. 
However, in order to theoretically estimate the plausible magnitudes of the non-tensorial SGWB, and to experimentally make statistical inferences on parameters of modified gravity models from detector data, 
it is necessary to consider the astrophysical origins of the non-tensorial SGWB.
%
Furthermore, obtaining astrophysicically motivated energy spectra may allow us to more efficiently search for the non-tensorial SGWB using a more optimal  matched filtering technique \cite{allen} --- than  simply assuming a power-law spectrum. 

In this paper, we focus on the SGWB in the transverse scalar mode of the Brans-Dicke (BD) theory: identifying its astrophysical origins, and obtaining its energy spectrum (as a function of the BD coupling constant).  Candidates for sources of the SGWB include gravitational stellar collapses and compact binary mergers.  As we will see, the existence of monopole scalar radiations makes stellar collapses by far the major contributor to this SGWB.  This differs significantly from the tensorial SGWB in GR, which is dominated by BBH and BNS mergers.

This paper is organized as follows: In Sec.~II, we will give an overview of the scalar GW in Brans-Dicke theory and its relation to scalar SGWB. In Secs.~III and IV, we will calculate the contributions to the scalar SGWB from compact binary coalescences, including BBH and BNS mergers, as well as the contribution from gravitational stellar collapses.   In Sec.~V, we will explore how the scalar SGWB depends on variations in the underlying population models.  In Sec.~VI, we will discuss the detectability and possible constraints on the BD coupling constant $\omega_{\rm BD}$ from current and future observations. Finally, in Sec.~VII,  we will draw conclusions and suggest prospective research directions.

\section{Scalar GW in Brans-Dicke Theory and Relation to SGWB}

In the Brans-Dicke (BD) theory, the Lagrangian density of the gravity sector in the original conformal frame is given by

\begin{align}
&\mathcal{L}_{\text{BD}}= \sqrt{-g} \left[ \phi R - \omega_\text{BD}\ \frac{\partial^\mu\phi\partial_\mu\phi \label{1}}{\phi} \right],
\end{align}
where the Ricci scalar $R$ is associated with the spacetime metric tensor $g_{\mu\nu}$. The scalar field $\phi$ is related to the gravitational constant $G$ via the relation 
\begin{equation} 
G \phi_0 = \frac{2\omega_\text{BD}+4}{2\omega_\text{BD}+3}
\end{equation}
where $\phi_0$ is the value of the scalar field at null infinity. The matter sector Lagrangian density remains the same as in GR, which means the scalar field does not couple to the matter fields directly. When the model parameter $\omega_\text{BD}$ approaches to infinity, Brans-Dicke theory recovers to GR.  In this rest of this section, we shall discuss the polarization content of GWs in the BD theory, and describe the energy content of the scalar SGWB. Details can be found in Refs.~\cite{brans, maggiore}.

\subsection{GWs in the BD Theory}

To study  GWs in Brans-Dicke theory, we perform a perturbation of the metric tensor and the scalar field around the Minkowski spacetime and the null infinity value, respectively:
\begin{align}
g_{\mu\nu} = \eta_{\mu\nu} + h_{\mu\nu}, \qquad \phi = \phi_0 + \delta \phi,
\end{align}
where components of the metric tensor in Minkowski spacetime is chosen to be $\eta_{\mu\nu} = \text{diag}(-1,1,1,1)$. 
The perturbative Lagrangian contains a quadratic cross term: $ h_{\mu\nu} (\partial^\mu \partial^\nu - \eta^{\mu\nu}\partial^2)\delta \phi$. To eliminate this term we redefine the following physical degrees of freedom:
\begin{align}
H_{\mu\nu} = h_{\mu\nu} + \eta_{\mu\nu}\frac{\delta \phi}{\phi_0}\,.
\end{align}
Under these treatments, the perturbative Lagrangian is expressed as:
\begin{align}
&\mathcal{L}_{\text{BD}}=\mathcal{L}_{\text{BD}}^\text{kin} + \mathcal{L}_{\text{BD}}^\text{S} + \mathcal{L}_{\text{BD}}^\text{other}\,, \nonumber \\
&\mathcal{L}_{\text{BD}}^\text{kin} =\frac{\phi_0}{2} H^{\mu\nu} V_{\mu\nu\rho\sigma}H^{\rho\sigma} + \frac{\omega_\text{BD}+3/2}{\phi_0}\ \eta^{\mu\nu} \delta\phi\ \partial_\mu \partial_\nu\delta\phi\,, \nonumber\\
&\mathcal{L}_{\text {BD}}^\text{S}=\frac{\omega_\text{BD} + 3/2}{\phi_0}\left(H^{\mu\nu}-\frac{1}{2}\eta^{\mu\nu} H\right) \partial_\mu\delta\phi\partial_\nu\delta\phi\ . \label{Lagrangians}
\end{align}
Here $\mathcal{L}_{\text{BD}}^\text{kin}$ represents the kinetic terms for the tensor field $H_{\mu\nu}$ and the scalar field $\delta\phi$, where the operator $V_{\mu\nu\rho\sigma}$ is defined as:
\begin{align}
V_{\mu\nu\rho\sigma}= \frac{1}{2}&\Big[(\eta_{\mu\rho}\eta_{\nu\sigma}-\eta_{\mu\nu}\eta_{\rho\sigma})\partial^2+ \eta_{\mu\nu}\partial_\rho\partial_\sigma \nonumber\\
&+\eta_{\rho\sigma}\partial_\mu\partial_\nu -\eta_{\mu\rho}\partial_\nu\partial_\sigma -\eta_{\nu\sigma}\partial_\mu\partial_\rho \Big] \ . \nonumber
\end{align}
The Lagrangian $\mathcal{L}_{\text{BD}}^\text{S}$ contains the leading interaction terms between the scalar and the tensor fields. Later we shall show that it relates to the scalar stress-energy tensor. The third term  $\mathcal{L}_{\text{BD}}^\text{other}$ contains other higher order interaction terms. Notice that the Lagrangians in Eq.~\eqref{Lagrangians} is invariant under the infinitesimal diffeomorphism transformation $x^\mu\rightarrow x'^\mu=x^\mu + \xi^\mu(x)$:
\begin{align}
 \begin{cases}
&H_{\mu\nu} \rightarrow H'_{\mu\nu}=H_{\mu\nu} - \partial_\mu\xi_\nu - \partial_\nu\xi_\mu     \\
\\
&\delta\phi \rightarrow \delta\phi'=\delta\phi \ .
 \end{cases} \label{gauge}
\end{align}
The vacuum field equation for $H_{\mu\nu}$ is obtained from $\delta  \mathcal{L}_{\text{BD}}^\text{kin}/\delta H_{\mu\nu}=0$, which gives $V^{\mu\nu\rho\sigma}H_{\rho\sigma}=0$. If we choose the harmonic coordinate condition, this equation is reduced to
\begin{align}
\begin{cases}
\partial^2 H_{\mu\nu} = 0 \\
\partial^\mu H_{\mu\nu} - \frac{1}{2}\partial_\nu H = 0 \ .
\end{cases}\label{fieldequation}
\end{align}
Notice that the vacuum field equation Eq.~\eqref{fieldequation} and the gauge transformation Eq.~\eqref{gauge} have the same form as in GR, hence we can similarly gauge away redundant degrees of freedom which leave only two physical ones. 

However, GW detectors respond directly to the change in the spacetime metric, i.e. $h_{\mu\nu}$, which depends both on $H_{\mu\nu}$ and $\delta\phi$. As a result, three physical degrees of freedom remain for $h_{\mu\nu}$ \cite{maggiore}. More specifically, within a spatial slice, let $\Sigma_\Omega$ be the 2-D plane perpendicular to the plain wave propagation direction $\hat {\bold \Omega}$, and let $\hat{\bold m}$, $\hat{\bold n}$ be two the orthogonal unit vectors in $\Sigma_\Omega$, then we can find a gauge in which the plain wave can be expanded as,
\begin{align}
h_{ij}(x) = h_+(x)e_{ij}^+ + h_\times(x) e_{ij}^+ + h_\text S(x)  e_{ij}^\text S.
\end{align} 
Here, the amplitudes are related to $H_{ij}$ and $\delta \phi$ via
\begin{align}
h_+=e_+^{ij}H^\text{TT}_{ij}/2, \quad h_\times=e_\times^{ij}H^\text{TT}_{ij}/2, \quad h_\text S=-\delta \phi/\phi_0,
\end{align}
here $H^\text{TT}_{ij}$ is the transverse-traceless part of $H_{ij}$ \cite{300years}. The polarization tensors are expressed as
\begin{align}
&e^+_{ij}=\hat{m}_i\hat{m}_j-\hat{n}_i\hat{n}_j, \qquad e^\times_{ij}=\hat{m}_i\hat{n}_j+\hat{n}_i\hat{m}_j,\nonumber \\
&e^S_{ij}=\hat{m}_i\hat{m}_j+\hat{n}_i\hat{n}_j. \label{polarizations}
\end{align}
Here $\bold e^{+,\times}$ and $\bold e^\text S$ represent tensor and scalar polarizations of GW respectively because under an $\mathrm{SO}(2)$ rotation in $\Sigma_\Omega$ plane: $\hat{\bold m}'+ i\hat{\bold n}'=\text{exp}(i\theta)(\hat{\bold m}+ i\hat{\bold n})$, they behave as ${\bold e^+}'+ i{\bold e^\times}'=\text{exp}(2i\theta)(\bold e^+ + i\bold e^\times)$ and ${\bold e^S}'=\bold e^S$. 

\subsection{The scalar SGWB }

We expect the presence of the scalar GW would give rise to a stochastic background with scalar polarization, which is described by the dimensionless energy density spectrum:
\begin{align}
\label{eqOmegaS}
\tilde\Omega_\text{S}(f)=\frac{1}{\rho_c} \frac{d\rho_\text S}{d\ln{f}}.
\end{align}
In this equation, $\rho_c = 3H_0^2/8\pi G$ is the critical density to close the Universe with $H_0$ representing the Hubble constant. The energy density of the scalar GW $\rho_\text S$ relates to the scalar stress-energy tensor $T_\text{S}^{\mu\nu}$ via $\rho_\text S = T_\text{S}^{00}$, with 
\begin{align}
T_\text{S}^{\mu\nu}=\frac{1}{8\pi}\frac{\partial \mathcal{L}_{\text {BD}}^\text{S}}{\partial H_{\mu\nu}}=\frac{\omega_\text{BD}+2}{8\pi G}\left(\partial^\mu h_\text S \partial^\nu h_\text S  - \frac{\eta^{\mu\nu}}{2}\partial^\rho h_\text S \partial_\rho h_\text S\right)\ . \label{TS}
\end{align}
Combining with the field equation $\partial^2 h_\text S=0$ and averaging over several wave length, we obtain \cite{maggiore}
 \begin{align}
 \label{eqrhoS}
 \rho_\text{S} = \frac{\omega_\text{BD}+2}{8\pi G}\left< \dot h_\text{S}^2(x)  \right>
 \end{align}

 Under the assumption that the stochastic background is stationary, isotropic and Gaussian, the ensemble average of the Fourier transformed amplitude $\tilde h_\text{S}(f,\hat\Omega)$  
 satisfies 
 \begin{align}
 \left<\tilde h^*_\text{S}(f,\hat\Omega) \tilde h_\text{S}(f',\hat\Omega') \right> =\frac{5}{8\pi} \delta(\hat\Omega - \hat\Omega')\delta(f-f') H_\text{S}(f).
 \end{align}
 Here  $\tilde h_\text{S}(f,\hat\Omega)$ relates to $h(t,\mathbf{x})$ via
 \begin{equation}
 h_S(t,\mathbf{x}) = \int \frac{d^3\mathbf{k}}{(2\pi)^3} e^{-i\omega(\mathbf{k}) t+i\mathbf{k} \cdot\mathbf{x}}\tilde h_S + \mbox{c.c.}
 \end{equation}
with $\omega(\mathbf{k}) = |\mathbf{k}|/c$, and 
 \begin{equation}
 \mathbf{k} =2\pi f \hat\Omega/c
 \end{equation}
 where $\hat \Omega$ is the unit vector along the direction of $\mathbf{k}$. The quantity  $H_\text S$ is defined as the spectral density for scalar GW. The factor of $5/8\pi$ follows the same convention in \cite{asearchfor, callister}.

Under this definition, $\tilde H_S$ is related to the scalar spectral density  $\tilde\Omega_\text{GW}^\text{S}$ [defined in Eq.~\eqref{eqOmegaS}--\eqref{eqrhoS}], via 
\begin{align}
\tilde\Omega_\text{S}(f) = (\omega_{BD}+2) \frac{20\pi^2}{3H_0^2}f^3 H_\text{S}(f) \ . \label{OmegaTildeHS}
\end{align}
As we shall see later in Sec.~\ref{sec:detectability}, the quantity $H_S$ is directly related to the detectability of the scalar SGWB [see Eq.~\eqref{SNR}].  In this way, even though $\tilde\Omega_S$ is directly proportional to the energy density of the scalar wave, detectability of the background, given the same $\tilde\Omega_S$, still depends on the BD coupling constant $\omega_{\rm BD}$ .  This is related to the violation of the Isaacson formula in BD theory~\cite{allen, isaacson}.  Instead, following the same convention as Ref.~\cite{asearchfor}, we define a new quantity 
\begin{align}
\Omega_\text{S}(f) = \frac{\tilde\Omega_\text{S}(f)}{\omega_{BD}+2}   = \frac{20\pi^2}{3H_0^2}f^3 H_\text{S}(f) \,,
\label{Omega}
\end{align}
In the following discussions, we will keep using this redefined energy density spectrum to describe the scalar SGWB.

\section{Scalar and Tensor SGWB from Mergers of Compact Binary System}

\subsection{Tensor SGWB from Compact Binary Mergers in BD Theory}

In GR, the SGWB has only tensor polarization and the major contribution within the bandwidth of ground based GW detectors is from the mergers of BBH, with $\Omega_\text{T}(f=25\text{Hz}) \simeq 1.1\times 10^{-9}$ \cite{gw150914implications}. Besides BBH, the mergers of BNS has a comparable contribution to the SGWB, with $\Omega_\text{T}(f=25\text{Hz}) \simeq 0.7\times 10^{-9}$ \cite{gw170817implications}.
In BD, we expect the tensor SGWB takes approximately  the same value as in GR, which is predicted from the relation \cite{brunetti, connell}:
\begin{align}
P_\text{T}^\text{(BD)} = \frac{2\omega_{\rm BD}+3}{2\omega_{\rm BD}+4} P_\text{T}^\text{(GR)},
\end{align}
where $P_\text{T}^\text{(BD)}$ and $P_\text{T}^\text{(GR)}$ denote the power emitted in GW with tensor polarization from a system of binary stars in BD and in GR respectively, at the same orbital frequency.
For a large $\omega_\text{BD}$, we expect the ratio between the two powers is approximately equal to one.   As will be shown later in the next section, in BD most  of gravitational radiation by binary stars is from tensor GW --- with scalar radiation suppressed by $\omega_{\rm BD}$. Consequently, the coalescing trajectory of the compact binary system which is mainly a result from GW radiation, as well as the spectrum of tensor GW radiation, is nearly unchanged as in GR.

\subsection{Scalar Radiation from a Compact Binary in BD Theory}

As for the scalar part, the story is completely different: the contribution to the scalar SGWB from the mergers of BBH is exactly zero. This is a direct implication from Hawking's {\it no scalar-hair} theorem of black holes in BD theory of gravity \cite{hawking}. The theorem states that for black holes in BD the exterior spacetine geometry is the same as in GR and the scalar field $\phi$ takes a constant value. Since $h_\text{S}=0$ everywhere, there is no scalar GW radiation from the merger of BBH. 

On the other hand, the no scalar-hair theorem does not forbid scalar GWs emitted from mergers of BNS. Within the bandwidth of ground based GW detectors, the background is mainly from the inspiral stage, since BNS merger frequency is above 2 kHz \cite{gw170817implications}. The power of scalar GW emission from inspiraling binary systems has been studied in \cite{brunetti, connell}. Contrary to the tensor case, the scalar GW has monopole and dipole radiations in addition to quadrupole radiation. In the limit of vanishing eccentricity $e\rightarrow 0$ (this assumption should be valid since the orbital angular momentum should have been radiated away from GW emission for coalescing binary systems as they enter the band of ground-based detectors) the scalar energy spectrum for the monopole radiation ($j=0$) is negligible
 (binary systems with circular orbit have no monopole moment), while the dipole ($j=1$) and quadruple ($j=2$) are given by
\begin{align}
&\frac{dE_\text{S}^{j=1}}{df}=\frac{1}{\omega_\text{BD}+2}\frac{5}{48}\left(\frac{\text{BE}_1}{m_1}-\frac{\text{BE}_2}{m_2}\right)\frac{m_1 m_2}{m_1+m_2}f^{-1} \nonumber \\
&\frac{dE_\text{S}^{j=2}}{df}(f)=\frac{1}{\omega_\text{BD}+2}\frac{(\pi G)^{\frac{2}{3}}}{36}\frac{m_1 m_2}{(m_1+m_2)^\frac{1}{3}}f^{-\frac{1}{3}} \ . \label{energy}
\end{align}
Here $f$ represents the frequency of GW, $m_1$ and $m_2$ the masses of the two neutron stars in the binary system. The energy spectrum is derived from the relation to the power: $d E_\text{S}^j/df = P_\text{S}^j/\dot f$, where we adopt the power of scalar GW emission $P_\text{S}^j$ calculated by Brunetti et al. \cite{brunetti}. In the limit of $e\rightarrow 0$, the orbital frequency $F$ and the GW frequency $f$ are related by $f=j F$ for $j=1,2$. The  rate of change of the orbital frequency due to GW emission is the same as in GR \cite{mtw}:
\begin{align}
\dot F = \frac{48 \pi^\frac{8}{3}G^\frac{5}{3}}{5} \frac{m_1 m_2}{(m_1+m_2)^\frac{1}{3}}(2F)^\frac{11}{3}
\end{align}
In Eq.~\eqref{energy}, $\text{BE}$ represents the binding energy of neutron stars and we adopt the model by Lattimer and Prakash \cite{lattimer}, which reads
\begin{align}
\frac{\text{BE}}{m} \simeq \frac{0.6 \ \beta}{1-0.5 \ \beta}, \label{BE}
\end{align}
where $\beta = G m/R$ with $R$ denoting the radius of the neutron star.
 
\subsection{Scalar SGWB from Compact Binaries in BD Theory}

The energy density spectrum of the produced SGWB can be obtained from the emission spectrum of a single BNS merger event via~\cite{phinney, gw150914implications}
\begin{align}
\Omega_\text{S}^{j}(f) = \frac{1}{\omega_\text{BD}+2}\frac{f}{\rho_c}\int_0^{z_\text{max}}dz \ \frac{R_m(z)\ \frac{dE^j_\text{S}}{df}(f_z)}{(1+z)H(z)} \ , \label{OmegadEdf}
\end{align}
where $f_z = (1+z)f$ is the frequency at emission. Note that the factor of $1/(\omega_\text{BD}+2)$ is from the definition of Eq.~\eqref{Omega}. Here we adopt the $\Lambda\text{CDM}$ cosmological model, with 
\begin{equation}
H(z) = H_0[\Omega_M(1+z)^3+\Omega_\Lambda]^{1/2},
\end{equation}
 where the Hubble constant $H_0=70 \text{km/s Mpc}$, $\Omega_M=0.3$ and $\Omega_\Lambda=0.7$. The redshift cutoff is chosen as $z_\text{max}=10$. 
 In Eq.~\eqref{OmegadEdf}, $R_m(z)$ is the BNS merger rate per comoving volume at redshift $z$. We adopt the same merger rate as in \cite{gw170817implications}, which is expressed as
\begin{align}
R_m(z) = R_m(0) \frac{\int_{t_\text{min}}^{t_\text{max}}dt_d\ R_f[z_f(z,t_d)] p(t_d)}{\int_{t_\text{min}}^{t_\text{max}}dt_d\ R_f[z_f(0,t_d)] p(t_d)} \ .
\end{align}
Here, $t_d$ denotes the time delay between formation and merger of BNS and $p(t_d)$ is its probability distribution function. We assume $p(t_d) \propto 1/t_d$ for $ t_\text{min} < t_d < t_\text{max}$, with $t_\text{min}=20\ \text{Mpc}$ and $t_\text{max}$ equal to the Hubble time $H_0^{-1}$.  The BNS formation rate $R_f(z)$ is assumed to be proportional to the star formation rate (SFR): $R_f(z)\propto \dot\rho_*(z)$. As in \cite{gw150914implications, gw170817implications} we adopt the GRB-based SFR model given in \cite{vangioni}, which is inferred from observed gamma-ray burst data at high redshift \cite{kistler}. The local BNS merger rate is inferred from GW170817 \cite{gw170817} with $R_m(0)= 1540\ \text{Gpc}^{-3} \text{yr}^{-1}$, and $z_f(z, t_d)$ is the redshift at the binary formation time $t_f = t(z)-t_d$, with $t(z)$ the age of the Universe at merger. 

In Eq.~\eqref{OmegadEdf}, the energy spectrum is given by Eq.~\eqref{energy} with the observed GW frequency $f$ replaced by the frequency at emission $f_z$. The frequency cutoff is at the innermost stable circular orbit (ISCO) \cite{zhu}: $f_\text{max} = f_\text{ISCO} \simeq 4400/(m_1+m_2)\ \text{Hz}$, with the mass in the unit of $M_\odot$. As in \cite{gw170817implications}, the neutron star masses $m_1$ and $m_2$ in the binary are assumed to follow a uniform distribution ranging from $1$ to $2\ M_\odot$. We adopt the neutron star mass-radius relation from the baseline model of Steiner et al. \cite{steiner}. Within our range of $m$, the radius is around $R\approx 12\ \text{km}$.

\begin{figure}[t]
\centering
\includegraphics[width=0.45\textwidth]{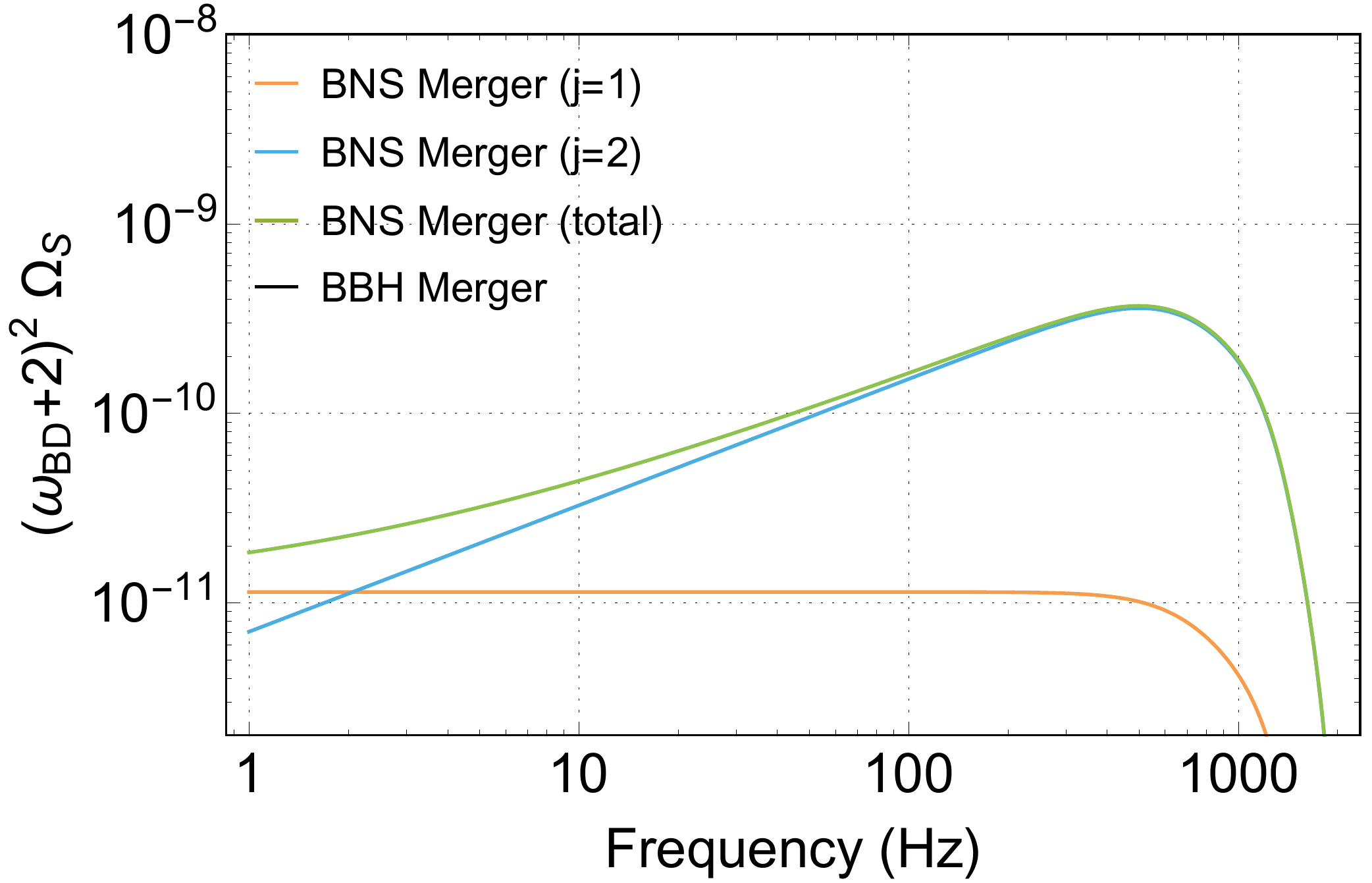}
\caption{The scalar SGWB from compact binary systems. The yellow curve is the contribution from mergers of BNS with $j=1$, at low frequencies it follows a power law of $f^0$. The blue curve is the contribution from mergers of BNS with $j=2$, at low frequencies it follows a power law of $f^{2/3}$. The green curve is the total BNS scalar SGWB. The mergers of BBH has no contribution to the scalar SGWB, which is a direct consequence of Hawking's no scalar-hair theorem \cite{hawking}.}
\label{fig1}
\end{figure}

We show the resulting scalar SGWB in Fig~\ref{fig1}. Note that the energy density spectrum $\Omega_{\rm S}$ we have chosen to use scales with the BD parameter as $\Omega_\text{S}\propto (\omega_\text{BD}+2)^{-2}$. For BNS mergers, we predict $(\omega_\text{BD}+2)^{2}\ \Omega_\text{S}^{j=1}(f=25\text{Hz})=1.1\times10^{-11}$ with a power law of $f^0$ at low frequencies and $(\omega_\text{BD}+2)^{2}\ \Omega_\text{S}^{j=2}(f=25\text{Hz})=6.0\times10^{-11}$ with a power law of $f^{2/3}$ at low frequencies. For $f>10\text{Hz}$, the dipole $(j=1)$ contribution to the scalar SGWB is much less than the quadrupole $(j=2)$, which is a consequence from the small asymmetry between the two neutron stars in the binary system. Also, as discussed earlier BBH has no contribution to the scalar SGWB.

\section{Scalar SGWB from Stellar Gravitational Core Collapse}

It is well known that massive stars end their lives through gravitational core collapse. In GR, stellar core collapses only contribute a minor fraction to the total SGWB. For example, Crocker et al. \cite{crocker} predict an SGWB from the black hole ringdown following the collapse with $\Omega_\text{T}(f=25\text{Hz}) \simeq 2\times 10^{-12}$, (Fig.~6 of \cite{crocker}, model 2 \& 3). In \cite{buonanno} Buonanno et al. predict the background from the neutrino burst associated with the core collapse with $\Omega_\text{T}(f=25\text{Hz}) \simeq 1\times 10^{-13}$, (Fig.~1 of \cite{buonanno}, the optimistic model). The small contribution to the SGWB given the greater event rate of stellar collapses compared to binary mergers can be explained by the fact that in GR, the tensor GWs are emitted through secondary effects of stellar collapse: only the small asymmetry in the collapse gives rise to a non-zero quadrupole moment.

\subsection{Scalar Emission from Gravitational Core Collapse} 

However, we expect a different picture in BD: the scalar GW emission starts from 
the monopole order, which indicates even the perfectly spherical collapses are able to emit scalar GW. Further, the progenitors of collapse are sources of the scalar field, with a monopole moment proportional to $m/(2\omega_\text{BD}+4)$ \cite{shibata}. As the progenitor collapses, this scalar monopole is radiated away. In this way, the scalar GW is dominated by the memory effect at low frequencies~\cite{du}: the scalar GW slumps from a nonzero initial value $h_\text{S}^\text{ini}$ to a zero final value if the collapse remnant is a black hole or a different nonzero final value if the remnant is a neutron star. The change in the amplitude of the scalar field is expressed as \cite{du}:
\begin{align}
\Delta h_{ij}=
\begin{cases}
\displaystyle -\frac{1}{\omega_\text{BD}+2}\frac{G\ m}{r}\ e^\text{S}_{ij} &\quad\text{black hole remnant}\\
\\
\displaystyle -\frac{1}{\omega_\text{BD}+2}\frac{G(m-m_\text{NS})}{r}\ e^\text{S}_{ij} &\quad\text{neutron star remnant}
\end{cases}. 
\end{align}
Here, $m$ and $m_\text{NS}$ represent the mass of the progenitor and the mass of the remnant neutron star, respectively.

As discussed in \cite{du}, for ground-based GW detectors, most of the detection band is dominated over by the memory as the ``zero-frequency limit". The resulting scalar energy spectrum from the memory effect is 
\begin{align}
\frac{dE_\text{S}}{df}= \frac{G\left[m-m_\text{NS}\Theta(M_\text{BH}-m)\right]^2}{\omega_\text{BD}+2}\Theta(m-M_C)\Theta(f_\text{cut}-f), \label{dEdfcollapse}
\end{align}
where $M_C$ is the minimum mass for the progenitor to end its life via core collapse and $M_\text{BH}$ is the mass threshold above which the final product from collapse is a black hole instead of a neutron star. As suggested in \cite{smartt}, we choose $M_C = 8 M_\odot$ and $M_\text{BH}=25M_\odot$. The NS mass is chosen as $M_\text{NS} = 1.4 M_\odot$. The cutoff frequency of the memory effect is $f_\text{cut}\simeq 1/\tau_c$, where the collapsing time is approximated from the Oppenheimer-Snyder model \cite{oppenheimer}:
\begin{align}
\tau_c \simeq Gm\frac{\pi}{\sqrt{8\beta^3(1-2\beta)}},
\label{collapsingTime}
\end{align}
where $\beta$ is the same as in Eq.~\eqref{BE}. Here we choose $\beta=0.1$ for the progenitor as in \cite{du, shibata}. \\

\subsection{Scalar SGWB from Core Collapse}

From the individual energy spectrum,  the total scalar SGWB energy density spectrum can be obtained using knowledge of collapse rates throughout the age of the universe \cite{crocker},
\begin{align}
\Omega_\text{S}(f) = \frac{1}{\omega_\text{BD}+2}\frac{f}{\rho_c}\int_0^{z_\text{max}}dz\int_{M_C}^{M_\text{max}}dm \ \frac{\frac{dR_c}{dm}(z,m)\ \frac{dE_\text{S}}{df}(f_z,m)}{(1+z)H(z)} \ . \label{OmegaCollapse}
\end{align}
 
\begin{figure}[t]
\centering
\includegraphics[width=0.45\textwidth]{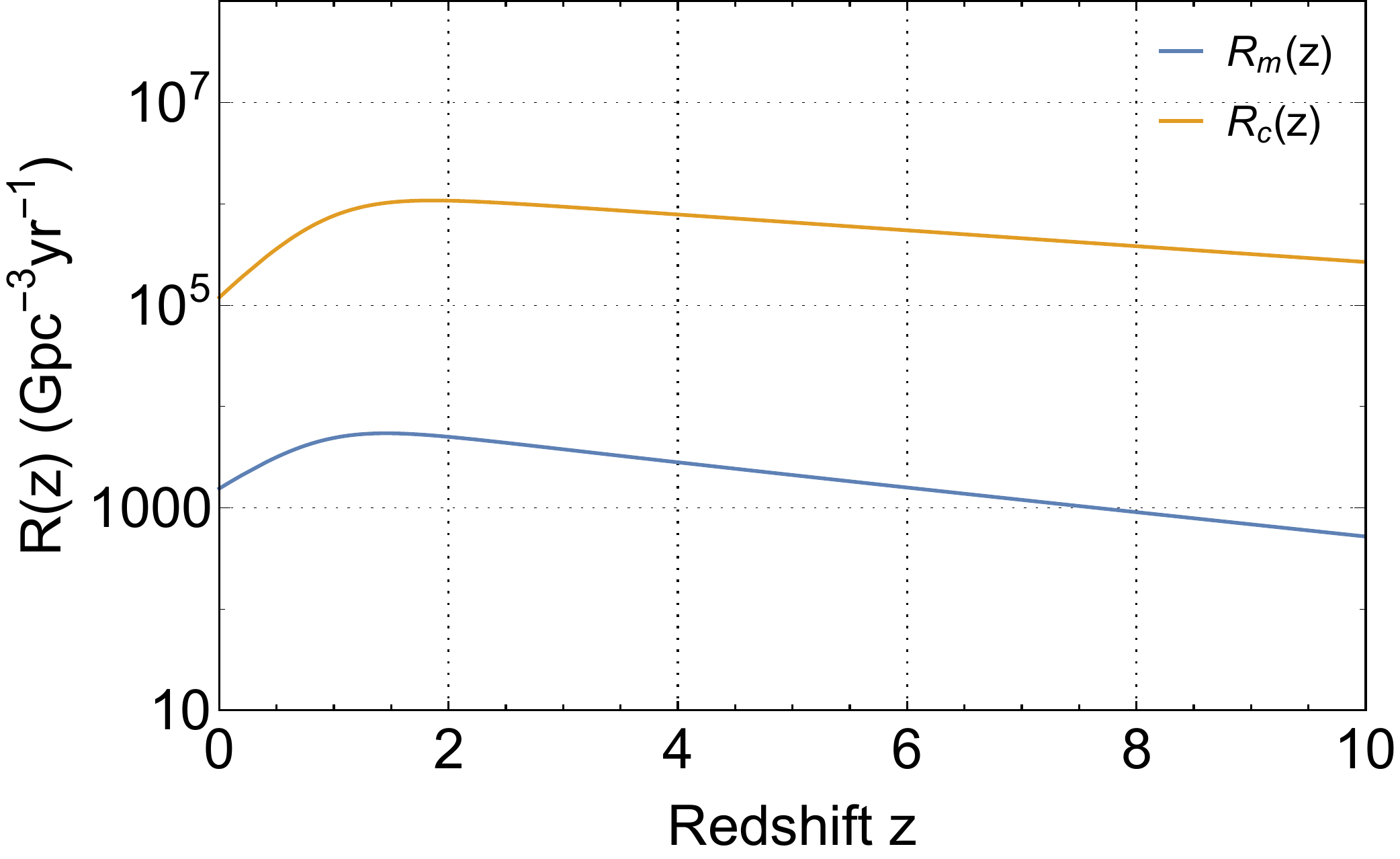}
\caption{Comparison between the BNS merger rate $R_m(z)$ and the core collapse rate $R_c(z)$.}
\label{fig2}
\end{figure}

In this equation, $M_\text{max}$ is the upper limit of the mass of massive stars and here we choose $M_\text{max}=100 M_\odot$ as reference. The energy spectrum is from Eq.~\eqref{dEdfcollapse} and the other parameters are the same as in Eq.~\eqref{OmegadEdf}. The collapse rate density $dR_c/dm$ (the number of collapses per unit proper time, per unit co-moving volume and per progenitor mass) is related to the Star Formation Rate (SFR) and initial mass function $\xi$ via \cite{ferrari, zhu2010}
\begin{align}
\frac{dR_c}{dm}(z,m) = \frac{\dot\rho_*(z) \xi(m)}{\int_{M_\text{min}}^{M_\text{max}}d\mu\ \mu\ \xi(\mu)}. \label{Rc}
\end{align}
Here we use the same SFR as in Section III, and choose the Salpeter IMF: $\xi(m)\propto m^{-2.35}$, with $M_\text{min}=0.1 M_\odot$ and $M_\text{max}=100 M_\odot$ \cite{vangioni}. The total merger $R_c(z)$ rate between $M_C$ and $M_\text{max}$ together with the BNS merger rate $R_m(z)$ are shown in Fig~\ref{fig2}.

\begin{figure}[t]
\centering
\includegraphics[width=0.45\textwidth]{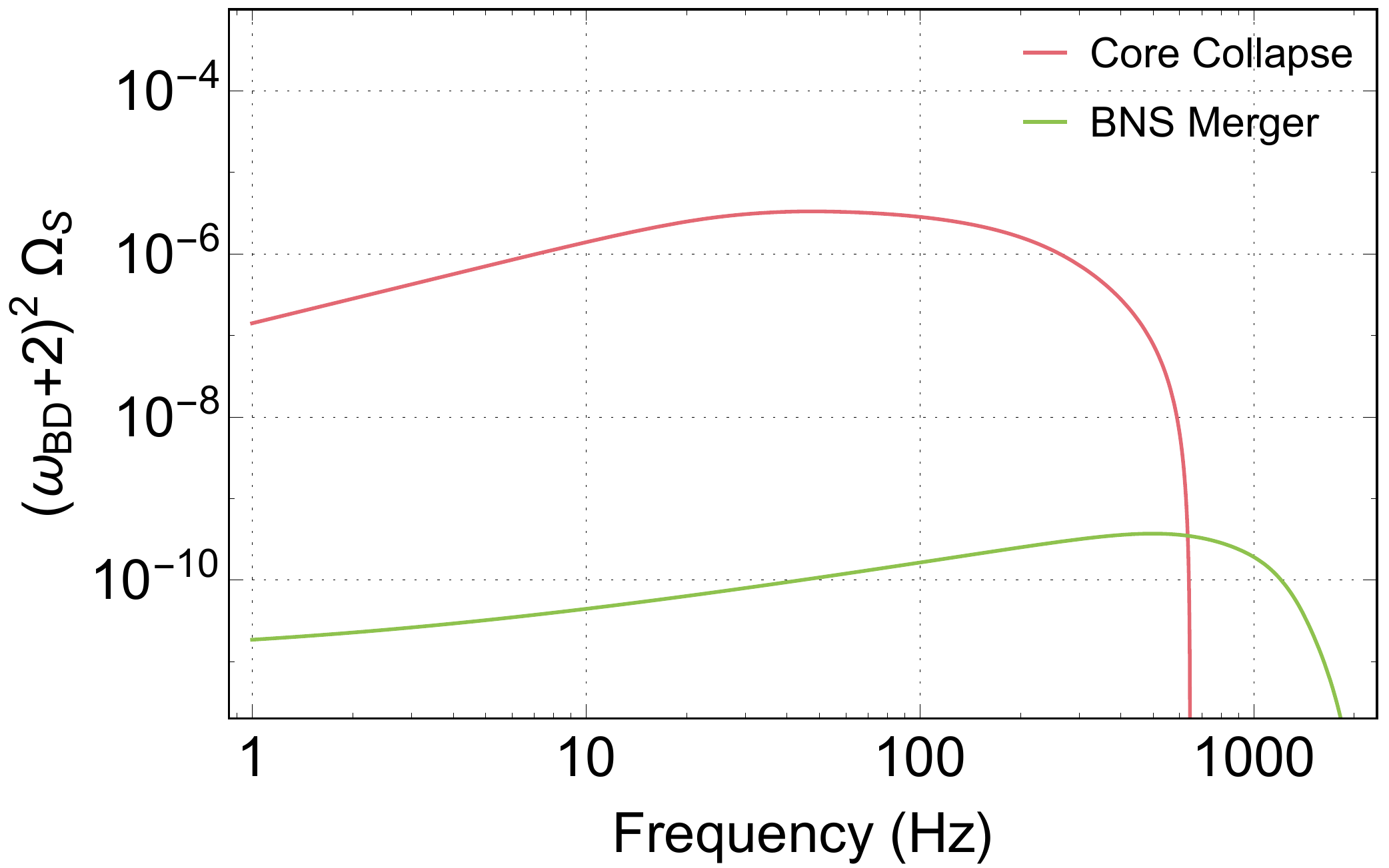}
\caption{Red curve: The scalar SGWB from core collapse. The model parameters are given in Section IV.
Green curve: The scalar SGWB from BNS merger, which is a sum of $j=1$ and $j=2$ radiation. The model parameters are given in Section III.}
\label{fig3}
\end{figure}

In Fig.~\ref{fig3}, we show the resulting scalar SGWB from core collapse. Same as the BNS scalar SGWB, the predicted energy density spectrum scales with the BD parameter as $\Omega_\text{S}\propto (\omega_\text{BD}+2)^{-2}$. At the reference frequency, $(\omega_\text{BD}+2)^{2}\ \Omega_\text{S}(f=25\text{Hz})=2.8\times10^{-6}$. At frequencies below $\sim 40\,$Hz,  $\Omega_S$ follows a power law of $f^\alpha$, with $\alpha=1$. Note that the core collapse scalar SGWB is around four orders of magnitude greater than BNS mergers. This difference can be accounted for by considering two factors. First, the collapse rate is much larger than the merger rate: at their peak values $R_c \simeq 10^6\ \text{Gpc}^-3\text{yr}^{-1}$ and $R_m \simeq 5\times10^3\ \text{Gpc}^-3\text{yr}^{-1}$. Second, the energy emitted from scalar GW radiation for a single collapse event is much larger than a merger event: notice that $E_\text{S}\propto m^2$, for mergers $m\sim1 M_\odot$ and for collapses $m\sim10 M_\odot$.

In Table \ref{tab1} we summarize the energy densities of SGWB with scalar polarization from varied sources, compared with the tensor SGWB.

\begin{table}[h]
\begin{tabular}{c|lc}
& $\Omega_T(f=25\,{\rm Hz})\quad  $ & $\quad(\omega_{\rm BD}+2)^2\Omega_S(f=25\,{\rm Hz})$ \\
\hline \hline
BBH  & $1.1\times 10^{-9}$~\cite{gw150914implications} & 0\\
BNS &  $0.7\times 10^{-9}$~\cite{gw170817implications} &  $7.1\times 10^{-11}$\\
Collapse &   $2\times 10^{-12}$~\cite{crocker}   & $2.8\times 10^{-6}$
\end{tabular}
\caption{Energy density of tensor and scalar SGWB at 25\,Hz, from various origins.} \label{tab1}
\end{table}

\subsection{Model dependence of the Core Collapse Scalar SGWB}
In this section we want to explore the influence to the core collapse scalar SGWB from alternative models.  From now on we refer the model described in Section IV as the {\it Baseline} model. More specifically, we consider four alternative models that follow.

(i) The {\it TimeDelay} model. In this model we take into account the time delay between the formation of a massive star and its core collapse. In this case, the collapse rate is modified as
\begin{align}
\frac{dR_c}{dm}(z,m) = \frac{\int_{t_\text{min}}^{t_\text{max}}dt_d \ \dot\rho_*[z_f(z,t_d)] \xi(m)p(t_d)}{\int_{M_\text{min}}^{M_\text{max}}d\mu\ \mu\ \xi(\mu)}.
\end{align}
With the other parameters the same as in Eq.~\eqref{Rc}, we assume the distribution as $p(t_d) = \delta(t_d - T(m))$, with $T(m)$ the lifetime of a star with mass $m$. In addition, we use the relation $T(m) = T_{\odot}(m/M_\odot)^{-2.5}$ for main sequence stars, with $M_\odot$ representing the solar mass and $T_\odot = 10^4\ \text{Myr}$. 

\begin{figure}[t]
\centering
\includegraphics[width=0.45\textwidth]{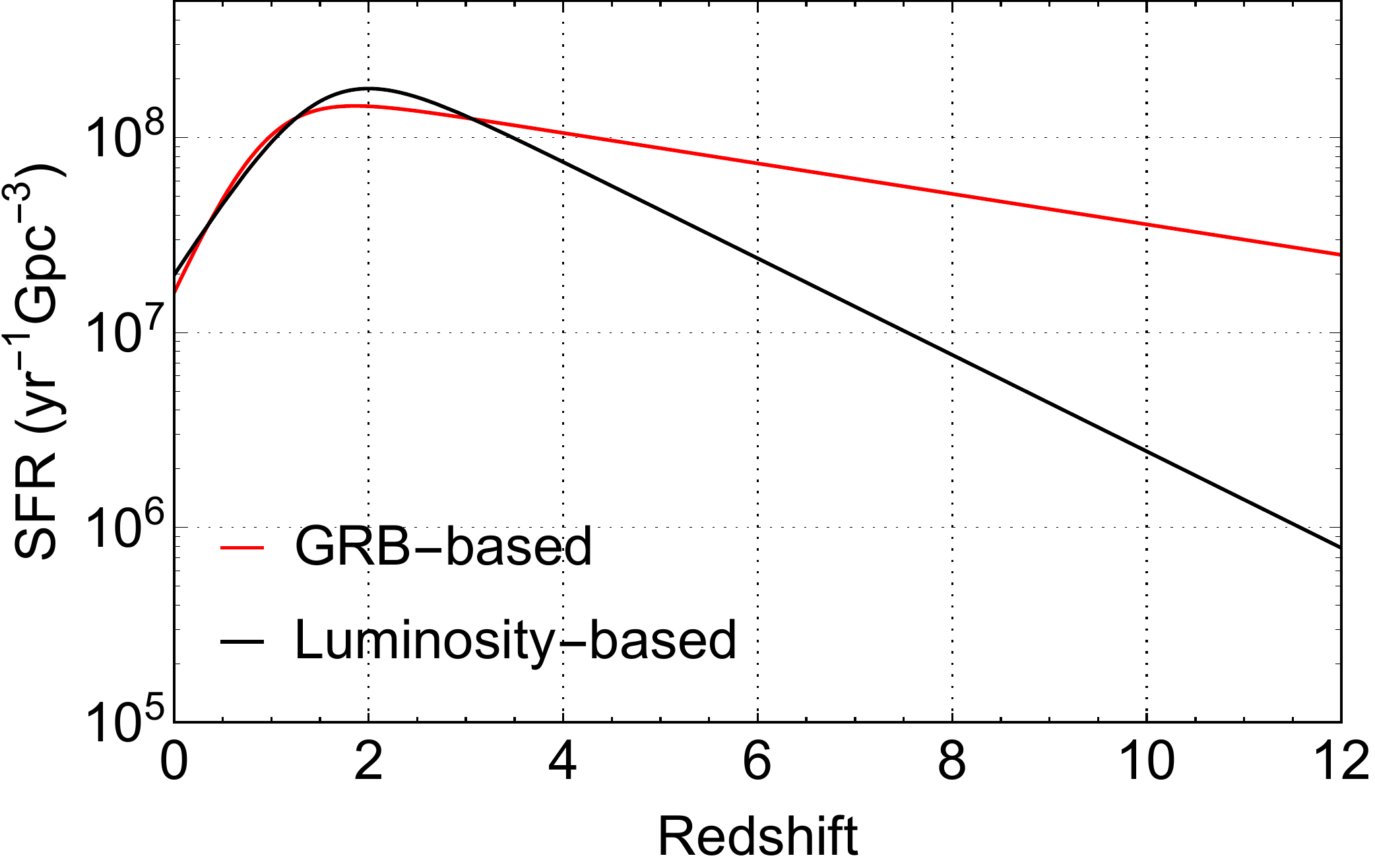}
\caption{Comparison between the GRB-based SFR and the Luminosity-based SFR.}
\label{fig4}
\end{figure}

(ii) The {\it AltSFR} model. In the {\it baseline} model we adopt an SFR model which is based on the GRB rate. In the {\it AltSFR} model we consider an alternative SFR model \cite{vangioni}, which based on the luminosity of star-forming galaxies \cite{behroozi}. This model is more conservative than the GRB-based SFR at high redshifts. We compare the two SFR models in Fig.~\ref{fig4}.

(iii) The {\it BHonly} model. In this model we only consider the scalar SGWB from core collapses into BHs. In this case the scalar energy spectrum is given by
\begin{align}
\frac{dE_\text{S}}{df}= \frac{G\ m^2}{\omega_\text{BD}+2}\Theta(m-M_\text{BH})\Theta(f_\text{cut}-f), 
\label{dEdfcollapseBHonly}
\end{align}
where the BH mass threshold $M_\text{BH}$ and the cutoff frequency  $f_\text{cut}$ are the same as in the {\it Baseline} model.

(iv) The {\it HighMass} model. To reflect the recent observations of massive stars with $M\sim 200-300M_\odot$ \cite{crowther}, we replace the mass upper limit $M_\text{max}$ to $200M_\odot$, with other parameters remaining the same.

\begin{figure}[t]
\centering
\includegraphics[width=0.45\textwidth]{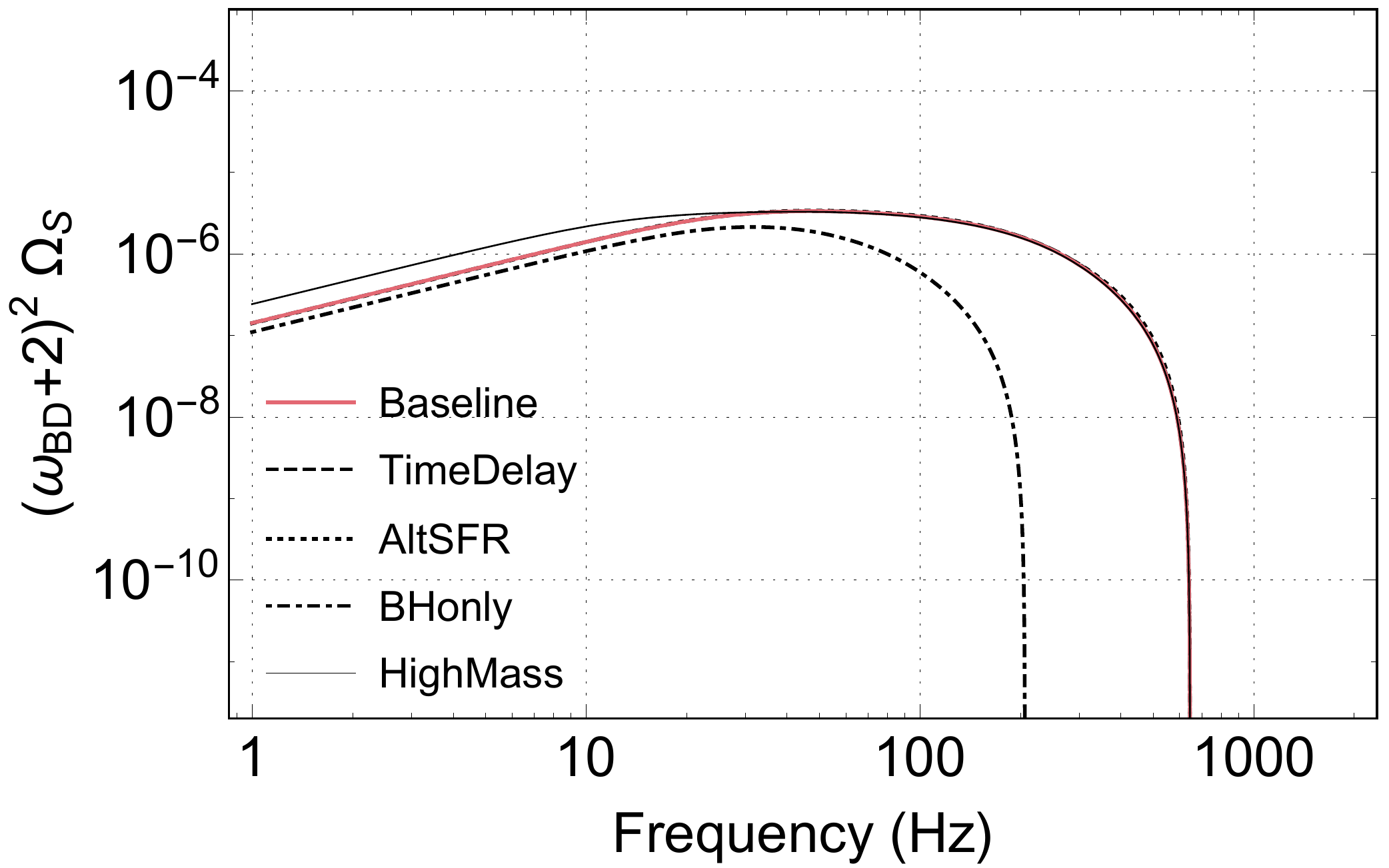}
\caption{The scalar overlap reduction function $\gamma_\text{S}$ for LIGO/Voyager and Einstein Telescope.}
\label{fig5}
\end{figure}

We show the scalar SGWB from the alternative models in Fig.~\ref{fig5}. We can see that {\it TimeDelay} and {\it AltSFR} have negligible influence on the background inside the detection band of ground-based detectors. 
The
{\it BHonly} and {\it HighMass} alter the background in low and high frequencies respectively. At the reference frequency $f = 25$~\text{Hz}, {\it HighMass} predicts a value of $\Omega_\text{S}$ that is $1.1$ times the {\it Baseline} value, while  {\it BHonly} predicts 0.7 the baseline value.  At this frequency, the impact from  {\it TimeDelay} and {\it AltSFR} to the scalar SGWB spectra is less than $1\%$.  At frequencies below $\sim 10\,$Hz, the {\it HighMass} model predicts somewhat higher $\Omega_S$, due to contributions from collapses of higher-mass objects.  At frequency $f > 100~\text{Hz}$, the only non-negligible change to the spectrum is from the {\it BHonly} model. This is because the stars which collapse into NSs have lower mass compared to those collapse into BHs. The cut-off frequency in Eq.~\ref{dEdfcollapseBHonly} is related to the collapsing time in Eq.~\ref{collapsingTime} which is shorter for collapsing stars with lower mass. As a result, the high frequency part of the spectrum is suppressed from the missing low mass progenitors.

\section{Detectability}
\label{sec:detectability}

Since the dominant contribution to the scalar SGWB in BD theory is from the core collapses, in this section we will focus on the scalar background predicted by the baseline core collapse model as described in Section IV. 

A resent analysis \cite{asearchfor} based on Advanced LIGO's first observing run (O1) has put the first upper limit on the scalar SGWB, with $\Omega_\text{S}(f=25\text{Hz})< 1.1\times10^{-7}$. Compared with our prediction $(\omega_\text{BD}+2)^{2}\ \Omega_\text{S}(f=25\text{Hz})=2.8\times10^{-6}$, it is straightforward to obtain $\omega_\text{BD}>3$. Much better upper limits are expected since the O1 data only includes an observation time of four months and the detectors are running below the designed sensitivity.

Next we want to explore the detectability from Advanced LIGO at its designed sensitivity and the planed future ground based GW detectors. The optimal signal-to-noise ratio (SNR) for the scalar SGWB between a pair of detectors is given by \cite{nishizawa, callister}
\begin{align}
\text{SNR}=\frac{3H_0^2}{10\pi^2}\sqrt{2T}\left(\int_0^\infty df\ \frac{\gamma_\text{S}(f)^2 \Omega_\text{S}(f)^2}{f^6 P_1(f)P_2(f)} \right)^{1/2} \label{SNR},
\end{align} 
where $P_{1,2}(f)$ are the detectors' noise spectral density and $\gamma_\text{S}(f)$ is the scalar overlap reduction function between the detectors \cite{maggiore}. Here we recall that it was the choice we had made in Eq.~\eqref{Omega} for  $\Omega_S$ that would lead to this expression for the  SNR, which is similar to  that for a tensor gravitational wave background.

\begin{figure}[t]
\centering
\includegraphics[width=0.45\textwidth]{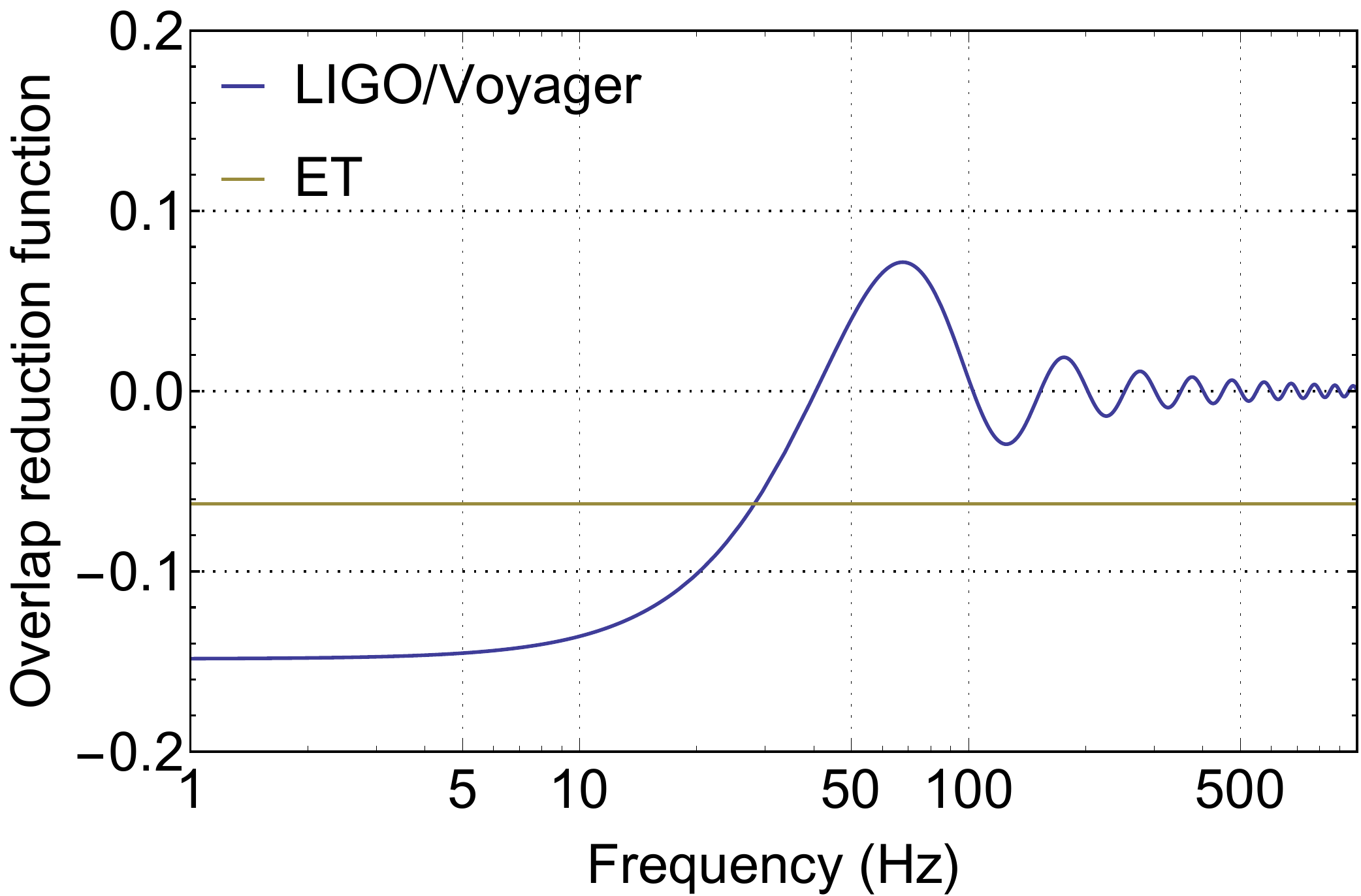}
\caption{The scalar overlap reduction function $\gamma_\text{S}$ for LIGO/Voyager and Einstein Telescope.}
\label{fig6}
\end{figure}

Here we consider the design sensitivity of Advanced LIGO \cite{ajith} and the planed sensitivities of LIGO Voyager \cite{voyager} and Einstein Telescope (ET) \cite{sathyaprakash}. The scalar overlap reduction function between the detectors at Hanford and Livingston is calculated in \cite{callister, nishizawa}, here we adopt the normalization convention as \cite{callister}. Voyager has the same overlap reduction function as LIGO. The co-located ET detectors have a constant $\gamma_\text{S}=-1/16$ for $f<1000 \text{Hz}$ (see Appendix for more details). These overlap reduction functions are shown in Fig.~\ref{fig6}. Note that our $\gamma_\text{S}$ for LIGO is one half of \cite{callister}, which is due to the fact that, as explained in Section II, only the breathing (traverse) and no longitudinal part of the scalar polarization exist in BD theory. 

\begin{figure}[t]
\centering
\includegraphics[width=0.45\textwidth]{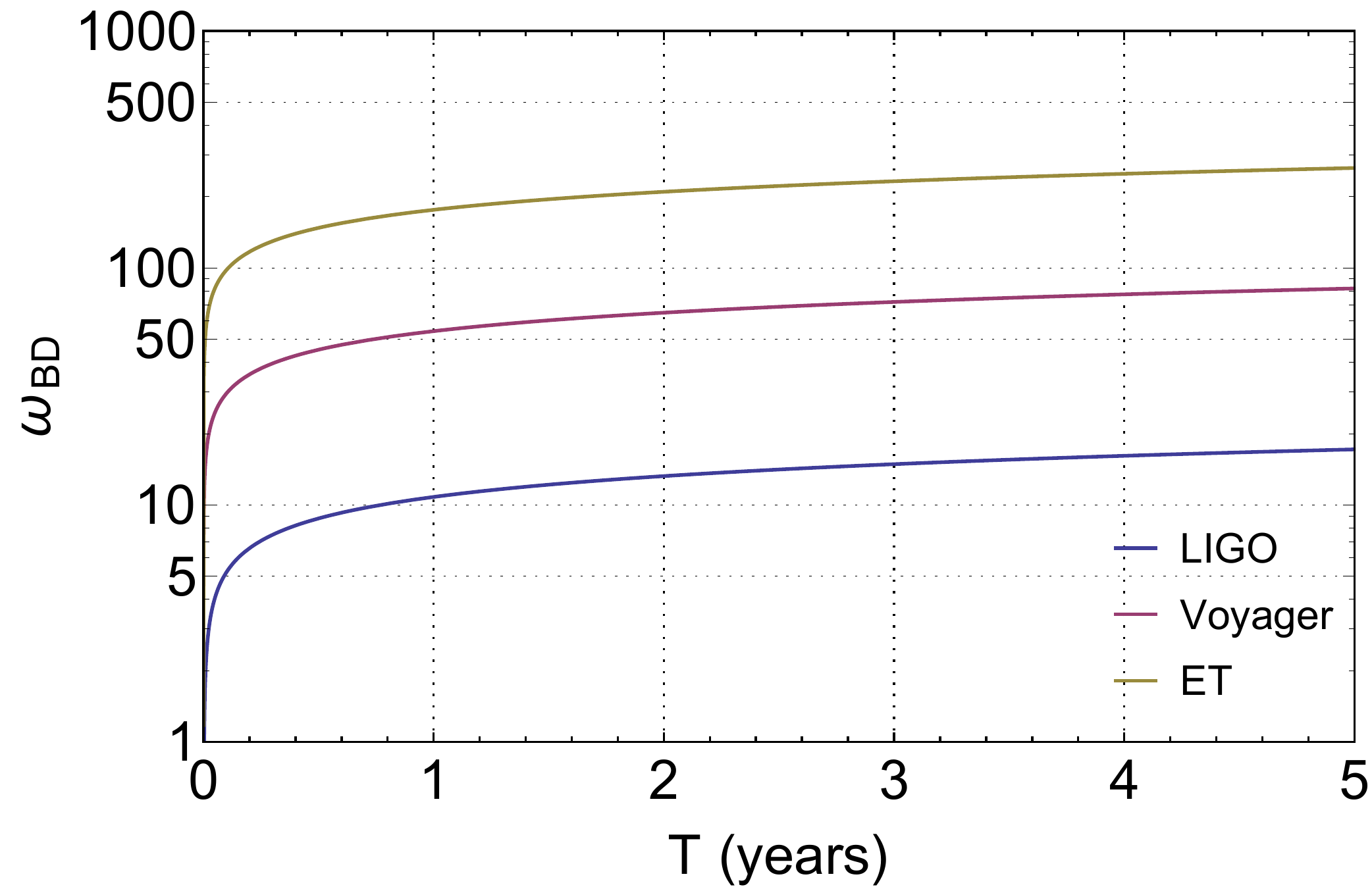}
\caption{Maximal detectable BD parameter $\omega_\text{BD}$ to reach an SNR threshold of $3$ from the scalar SGWB as a function of observation of time.}
\label{fig7}
\end{figure}

We show the maximal detectable $\omega_\text{BD}$ for LIGO, Voyager and ET to reach an SNR threshold of 3 in Table \ref{tab2} with observation times of 1 year and 5 years --- and in Fig.~\ref{fig7} for a range of observation times. With 5-year integration, to reach $\text{SNR}>3$ at ET, the BD parameter should be no less than $264$. On the other hand, the current cosmological constraints on BD set $\omega_\text{BD} > 692$ \cite{avilez} and the solar system data from the Cassini mission put a stronger constraint that $\omega_\text{BD} > 40000$ \cite{bertotti, will}. 

\begin{table}[ht] 
\begin{tabular}{ccccc} 
\hline \hline 
$T$ & LIGO  & Voyager & ET \\ 
\hline
1 yr &  $10.8$ &  $54.1$  &  $175.8$ \\ 
5 yrs & $17.1$ & $81.8$  &  $263.8$ \\ 
\hline \hline 
\end{tabular}
\caption{Maximal detectable BD parameter $\omega_\text{BD}$ to reach an SNR threshold of $3$ from the scalar SGWB with observation times of 1 year and 5 years.} \label{tab2} 
\end{table}

\section{Conclusions and Discussions}

In this paper, we studied the scalar SGWB in BD theory, from astrophysical sources, in particular compact binary mergers and stellar collapses.  Unlike the tensor SGWB in GR, we found that the scalar SGWB in the BD theory is dominated by stellar collapses, by roughly 4 orders of magnitude, over compact binary mergers. We have attributed this dominance to the higher rate of gravitational collapses than binary mergers, as well as the fact that scalar radiation does not require asymmetry. 

Furthermore, scalar radiation from stellar collapses, in the LIGO band, is mainly dominated by the memory wave --- as pointed out in an earlier paper ~\cite{du}.   Since the memory wave has a simple frequency dependence of $h (f) \sim 1/f$, this has lead to a $\Omega_S(f) \propto f$, which differs from the tenor SGWB, which as $\Omega_{\rm T} \propto f^{2/3}$.

For the dominant stellar-collapse scalar SGWB, we have studied a range of models, which had lead to consistent predictions, with the most significant uncertainty lying at low frequencies: up to within 30\% at $f=25~\text{Hz}$, mainly due to possible existence of heavier stars and the exclusion of the collapses whose remnant are NSs.

Upon obtaining the SGWB spectrum, we have estimated the detectability for current and  future detector networks.  It is estimated that 3rd-generation ground-based detectors can pose upper limit for $\omega_{\rm BD}$ around $\sim 300$.  

The potential bound for $\omega_{\rm BD}$ from our calculation is low compare with solar-system bounds, and somewhat lower than cosmological bounds, this nevertheless provides an  {\it independent} test.  More importantly,  having established that the scalar SGWB mainly arise from stellar collapses, we can further investigate other models that lead to scalar radiations, e.g., scalar-tensor theories in which $\omega(\phi)$ depends on the value of $\phi$ instead of being a constant. As we had shown in Ref.~\cite{du}, in such models the scalar memory, which dominates scalar radiation during collapse, can be significantly enhanced by such dependencies through {\it scalariation}~\cite{damour}, therefore might lead to much stronger SGWB enhanced by several orders of magnitude ~\cite{novak, harada}. In that case, we expect a considerable increase in the detectability from the current and the next generation of detectors.  We leave these for further studies. 

\section{Acknowledgements}
The author thanks Yanbei Chen for valuable discussions and the comments on the manuscript. The author is also grateful to Xi-Long Fan, Xiang-Cheng Ma and Atsushi Nishizawa for discussions.  We acknowledge support from the Brinson Foundation and the Simons Foundation.  Our research has also been supported by the National Science Foundation, through grants PHY-1404569, PHY-1708212 and PHY-1708213.

\section{Appendix: Scalar Overlap Reduction Function For Einstein Telescope}

\begin{figure}[t]
\centering
\includegraphics[width=0.45\textwidth]{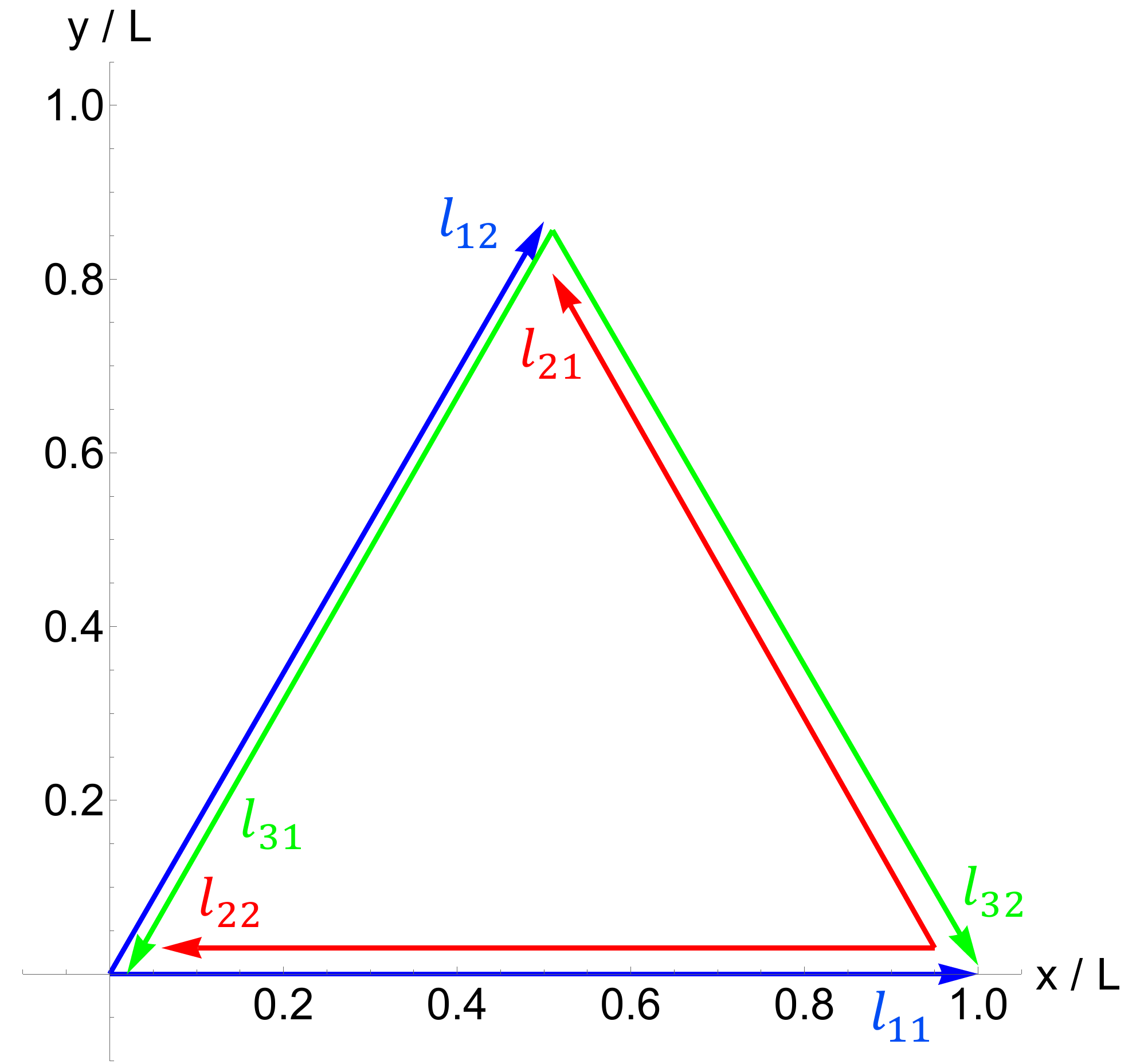}
\caption{Configuration of Einstein Telescope.}
\label{fig8}
\end{figure}

In this appendix we calculate the scalar overlap reduction function $\gamma_\text{S}(f)$ for Einstein Telescope (ET). The configuration of ET is shown in Fig.~\ref{fig8}. The coordinate system for the detectors are
\begin{align}
\begin{cases}
 \hat{\bold x} = (1,0,0) \\
 \hat{\bold y} = (0,1,0) \\
 \hat{\bold z} = (0,0,1) 
\end{cases}
\end{align}
In this coordinate system, the unit vectors for the ET detector arms are 
\begin{align}
&\hat{\bold l}_{11} = \hat{\bold x}, \qquad \hat{\bold l}_{12} = \frac{1}{2}\hat{\bold x}+\frac{\sqrt{3}}{2}\hat{\bold y},\nonumber \\
&\hat{\bold l}_{21} = -\frac{1}{2}\hat{\bold x} +\frac{\sqrt{3}}{2}\hat{\bold y}, \qquad \hat{\bold l}_{22} = - \hat{\bold l}_{11}, \nonumber\\
&\hat{\bold l}_{31} = -\hat{\bold l}_{12}, \qquad \hat{\bold l}_{32} = - \hat{\bold l}_{21}, 
\end{align}
The detector tensors for ET are express by
\begin{align}
D^{ij}_a=\frac{1}{2}(\hat l_{a1}^i \hat l_{a1}^j - \hat l_{a2}^i \hat l_{a2}^j),
\end{align}
where $a = 1,2,3$.

Suppose the GW is propagating along the angle $(\theta,\phi)$, the GW coordinate system can be constructed as
\begin{align}
\begin{cases}
 &\hat{\bold m} = \cos\theta\cos\phi\  \hat{\bold x} + \cos\theta\sin\phi\  \hat{\bold y}  -\sin\theta\  \hat{\bold z}    \\
 &\hat{\bold n} = -\sin\phi\  \hat{\bold x} + \cos\phi\  \hat{\bold y} \\
 &\hat{\bold \Omega} = \sin\theta\cos\phi\  \hat{\bold x} + \sin\theta\sin\phi\  \hat{\bold y}  + \cos\theta\  \hat{\bold z} 
\end{cases}
\end{align}

Then the angular pattern functions for scalar polarization are \cite{allen, nishizawa}
\begin{align}
F^\text{S}_a(\hat{\bold \Omega}) = \sum_{ij} D_a^{ij} e_\text{S}^{ij},
\end{align}
where the scalar polarization tensor $e_\text{S}^{ij}$ are given in Eq.~\eqref{polarizations}. It is straightforward to find that
\begin{align}
& F^\text{S}_1(\hat{\bold \Omega})  = \frac{1}{8} \sin^2{\theta}(-3 \cos{2 \phi} + \sqrt{3} \sin{2 \phi}) \nonumber\\
& F^\text{S}_2(\hat{\bold \Omega})  = \frac{1}{8} \sin^2{\theta}(3 \cos{2 \phi} + \sqrt{3} \sin{2 \phi}) \nonumber\\
& F^\text{S}_3(\hat{\bold \Omega})  = -\frac{\sqrt 3}{4} \sin^2{\theta}\sin{2\phi} 
\end{align}
The scalar overlap reduction function is defined as \cite{nishizawa, callister}
\begin{align}
\gamma_{ab}^\text{S}(f)=\frac{5}{8\pi}\int d\hat{\bold\Omega}\ e^{2\pi i f \bold{\hat\Omega}\cdot\Delta\bold x} F^\text{S}_a(\hat{\bold\Omega})F^\text{S}_b(\hat{\bold\Omega}).
\end{align}
Here we use the same normalization as \cite{callister}. For ET, the separation $|\Delta \bold x|$ is equal to the arm length $d=10\ \text{km}$. Hence, for $f < 10^3\ \text{Hz}$, the exponential function $e^{2\pi i f \bold{\hat\Omega}\cdot\Delta\bold x} \simeq 1$. In this case, 
\begin{align}
\gamma^\text{S}_{12}=\frac{5}{8\pi}\int_0^\pi d\theta\ \int_0^{2\pi}d\phi \left[-\frac{3}{64}(1+2\cos{4\phi})\sin^5{\theta}\right]=-\frac{1}{16}.
\end{align}
Similarly, we can show $\gamma^\text{S}_{23}=\gamma^\text{S}_{31}=-1/16$ for $f < 10^3\ \text{Hz}$.

\newpage
\bibliography{references}

\begin{thebibliography}{48}%
\makeatletter
\providecommand \@ifxundefined [1]{%
 \@ifx{#1\undefined}
}%
\providecommand \@ifnum [1]{%
 \ifnum #1\expandafter \@firstoftwo
 \else \expandafter \@secondoftwo
 \fi
}%
\providecommand \@ifx [1]{%
 \ifx #1\expandafter \@firstoftwo
 \else \expandafter \@secondoftwo
 \fi
}%
\providecommand \natexlab [1]{#1}%
\providecommand \enquote  [1]{``#1''}%
\providecommand \bibnamefont  [1]{#1}%
\providecommand \bibfnamefont [1]{#1}%
\providecommand \citenamefont [1]{#1}%
\providecommand \href@noop [0]{\@secondoftwo}%
\providecommand \href [0]{\begingroup \@sanitize@url \@href}%
\providecommand \@href[1]{\@@startlink{#1}\@@href}%
\providecommand \@@href[1]{\endgroup#1\@@endlink}%
\providecommand \@sanitize@url [0]{\catcode `\\12\catcode `\$12\catcode
  `\&12\catcode `\#12\catcode `\^12\catcode `\_12\catcode `\%12\relax}%
\providecommand \@@startlink[1]{}%
\providecommand \@@endlink[0]{}%
\providecommand \url  [0]{\begingroup\@sanitize@url \@url }%
\providecommand \@url [1]{\endgroup\@href {#1}{\urlprefix }}%
\providecommand \urlprefix  [0]{URL }%
\providecommand \Eprint [0]{\href }%
\providecommand \doibase [0]{http://dx.doi.org/}%
\providecommand \selectlanguage [0]{\@gobble}%
\providecommand \bibinfo  [0]{\@secondoftwo}%
\providecommand \bibfield  [0]{\@secondoftwo}%
\providecommand \translation [1]{[#1]}%
\providecommand \BibitemOpen [0]{}%
\providecommand \bibitemStop [0]{}%
\providecommand \bibitemNoStop [0]{.\EOS\space}%
\providecommand \EOS [0]{\spacefactor3000\relax}%
\providecommand \BibitemShut  [1]{\csname bibitem#1\endcsname}%
\let\auto@bib@innerbib\@empty
\bibitem [{\citenamefont {Abbott}\ \emph
  {et~al.}(2016{\natexlab{a}})\citenamefont {Abbott} \emph
  {et~al.}}]{gw150914}%
  \BibitemOpen
  \bibfield  {author} {\bibinfo {author} {\bibfnamefont {B.~P.}\ \bibnamefont
  {Abbott}} \emph {et~al.},\ }\href@noop {} {\bibfield  {journal} {\bibinfo
  {journal} {Physical review letters}\ }\textbf {\bibinfo {volume} {116}},\
  \bibinfo {pages} {061102} (\bibinfo {year} {2016}{\natexlab{a}})}\BibitemShut
  {NoStop}%
\bibitem [{\citenamefont {Abbott}\ \emph
  {et~al.}(2016{\natexlab{b}})\citenamefont {Abbott} \emph
  {et~al.}}]{gw151226}%
  \BibitemOpen
  \bibfield  {author} {\bibinfo {author} {\bibfnamefont {B.~P.}\ \bibnamefont
  {Abbott}} \emph {et~al.},\ }\href@noop {} {\bibfield  {journal} {\bibinfo
  {journal} {Phys Rev Lett}\ }\textbf {\bibinfo {volume} {116}},\ \bibinfo
  {pages} {241103} (\bibinfo {year} {2016}{\natexlab{b}})}\BibitemShut
  {NoStop}%
\bibitem [{\citenamefont {Abbott}\ \emph
  {et~al.}(2017{\natexlab{a}})\citenamefont {Abbott} \emph
  {et~al.}}]{gw170104}%
  \BibitemOpen
  \bibfield  {author} {\bibinfo {author} {\bibfnamefont {B.~P.}\ \bibnamefont
  {Abbott}} \emph {et~al.},\ }\href@noop {} {\bibfield  {journal} {\bibinfo
  {journal} {Physical Review Letters}\ }\textbf {\bibinfo {volume} {118}},\
  \bibinfo {pages} {221101} (\bibinfo {year} {2017}{\natexlab{a}})}\BibitemShut
  {NoStop}%
\bibitem [{\citenamefont {Abbott}\ \emph
  {et~al.}(2017{\natexlab{b}})\citenamefont {Abbott} \emph
  {et~al.}}]{gw170814}%
  \BibitemOpen
  \bibfield  {author} {\bibinfo {author} {\bibfnamefont {B.~P.}\ \bibnamefont
  {Abbott}} \emph {et~al.},\ }\href@noop {} {\bibfield  {journal} {\bibinfo
  {journal} {Physical review letters}\ }\textbf {\bibinfo {volume} {119}},\
  \bibinfo {pages} {141101} (\bibinfo {year} {2017}{\natexlab{b}})}\BibitemShut
  {NoStop}%
\bibitem [{\citenamefont {Abbott}\ \emph
  {et~al.}(2017{\natexlab{c}})\citenamefont {Abbott} \emph
  {et~al.}}]{gw170817}%
  \BibitemOpen
  \bibfield  {author} {\bibinfo {author} {\bibfnamefont {B.~P.}\ \bibnamefont
  {Abbott}} \emph {et~al.},\ }\href@noop {} {\bibfield  {journal} {\bibinfo
  {journal} {Physical Review Letters}\ }\textbf {\bibinfo {volume} {119}},\
  \bibinfo {pages} {161101} (\bibinfo {year} {2017}{\natexlab{c}})}\BibitemShut
  {NoStop}%
\bibitem [{\citenamefont {Abbott}\ \emph
  {et~al.}(2016{\natexlab{c}})\citenamefont {Abbott} \emph
  {et~al.}}]{gw150914implications}%
  \BibitemOpen
  \bibfield  {author} {\bibinfo {author} {\bibfnamefont {B.}~\bibnamefont
  {Abbott}} \emph {et~al.},\ }\href@noop {} {\bibfield  {journal} {\bibinfo
  {journal} {Physical review letters}\ }\textbf {\bibinfo {volume} {116}},\
  \bibinfo {pages} {131102} (\bibinfo {year} {2016}{\natexlab{c}})}\BibitemShut
  {NoStop}%
\bibitem [{\citenamefont {Abbott}\ \emph
  {et~al.}(2018{\natexlab{a}})\citenamefont {Abbott} \emph
  {et~al.}}]{gw170817implications}%
  \BibitemOpen
  \bibfield  {author} {\bibinfo {author} {\bibfnamefont {B.~P.}\ \bibnamefont
  {Abbott}} \emph {et~al.},\ }\href@noop {} {\bibfield  {journal} {\bibinfo
  {journal} {Physical review letters}\ }\textbf {\bibinfo {volume} {120}},\
  \bibinfo {pages} {091101} (\bibinfo {year} {2018}{\natexlab{a}})}\BibitemShut
  {NoStop}%
\bibitem [{\citenamefont {Abbott}\ \emph
  {et~al.}(2016{\natexlab{d}})\citenamefont {Abbott} \emph
  {et~al.}}]{gw150914test}%
  \BibitemOpen
  \bibfield  {author} {\bibinfo {author} {\bibfnamefont {B.~P.}\ \bibnamefont
  {Abbott}} \emph {et~al.},\ }\href@noop {} {\bibfield  {journal} {\bibinfo
  {journal} {Physical Review Letters}\ }\textbf {\bibinfo {volume} {116}},\
  \bibinfo {pages} {221101} (\bibinfo {year} {2016}{\natexlab{d}})}\BibitemShut
  {NoStop}%
\bibitem [{\citenamefont {Abbott}\ \emph
  {et~al.}(2017{\natexlab{d}})\citenamefont {Abbott} \emph
  {et~al.}}]{gw170817grb170817a}%
  \BibitemOpen
  \bibfield  {author} {\bibinfo {author} {\bibfnamefont {B.~P.}\ \bibnamefont
  {Abbott}} \emph {et~al.},\ }\href@noop {} {\bibfield  {journal} {\bibinfo
  {journal} {The Astrophysical Journal Letters}\ }\textbf {\bibinfo {volume}
  {848}},\ \bibinfo {pages} {L13} (\bibinfo {year}
  {2017}{\natexlab{d}})}\BibitemShut {NoStop}%
\bibitem [{\citenamefont {Eardley}\ \emph
  {et~al.}(1973{\natexlab{a}})\citenamefont {Eardley}, \citenamefont {Lee},
  \citenamefont {Lightman}, \citenamefont {Wagoner},\ and\ \citenamefont
  {Will}}]{eardley1}%
  \BibitemOpen
  \bibfield  {author} {\bibinfo {author} {\bibfnamefont {D.~M.}\ \bibnamefont
  {Eardley}}, \bibinfo {author} {\bibfnamefont {D.~L.}\ \bibnamefont {Lee}},
  \bibinfo {author} {\bibfnamefont {A.~P.}\ \bibnamefont {Lightman}}, \bibinfo
  {author} {\bibfnamefont {R.~V.}\ \bibnamefont {Wagoner}}, \ and\ \bibinfo
  {author} {\bibfnamefont {C.~M.}\ \bibnamefont {Will}},\ }\href@noop {}
  {\bibfield  {journal} {\bibinfo  {journal} {Physical Review Letters}\
  }\textbf {\bibinfo {volume} {30}},\ \bibinfo {pages} {884} (\bibinfo {year}
  {1973}{\natexlab{a}})}\BibitemShut {NoStop}%
\bibitem [{\citenamefont {Eardley}\ \emph
  {et~al.}(1973{\natexlab{b}})\citenamefont {Eardley}, \citenamefont {Lee},\
  and\ \citenamefont {Lightman}}]{eardley2}%
  \BibitemOpen
  \bibfield  {author} {\bibinfo {author} {\bibfnamefont {D.~M.}\ \bibnamefont
  {Eardley}}, \bibinfo {author} {\bibfnamefont {D.~L.}\ \bibnamefont {Lee}}, \
  and\ \bibinfo {author} {\bibfnamefont {A.~P.}\ \bibnamefont {Lightman}},\
  }\href@noop {} {\bibfield  {journal} {\bibinfo  {journal} {Physical Review
  D}\ }\textbf {\bibinfo {volume} {8}},\ \bibinfo {pages} {3308} (\bibinfo
  {year} {1973}{\natexlab{b}})}\BibitemShut {NoStop}%
\bibitem [{\citenamefont {Brans}\ and\ \citenamefont {Dicke}(1961)}]{brans}%
  \BibitemOpen
  \bibfield  {author} {\bibinfo {author} {\bibfnamefont {C.}~\bibnamefont
  {Brans}}\ and\ \bibinfo {author} {\bibfnamefont {R.~H.}\ \bibnamefont
  {Dicke}},\ }\href@noop {} {\bibfield  {journal} {\bibinfo  {journal}
  {Physical review}\ }\textbf {\bibinfo {volume} {124}},\ \bibinfo {pages}
  {925} (\bibinfo {year} {1961})}\BibitemShut {NoStop}%
\bibitem [{\citenamefont {Maggiore}\ and\ \citenamefont
  {Nicolis}(2000)}]{maggiore}%
  \BibitemOpen
  \bibfield  {author} {\bibinfo {author} {\bibfnamefont {M.}~\bibnamefont
  {Maggiore}}\ and\ \bibinfo {author} {\bibfnamefont {A.}~\bibnamefont
  {Nicolis}},\ }\href@noop {} {\bibfield  {journal} {\bibinfo  {journal}
  {Physical Review D}\ }\textbf {\bibinfo {volume} {62}},\ \bibinfo {pages}
  {024004} (\bibinfo {year} {2000})}\BibitemShut {NoStop}%
\bibitem [{\citenamefont {Nishizawa}\ \emph {et~al.}(2009)\citenamefont
  {Nishizawa}, \citenamefont {Taruya}, \citenamefont {Hayama}, \citenamefont
  {Kawamura},\ and\ \citenamefont {Sakagami}}]{nishizawa}%
  \BibitemOpen
  \bibfield  {author} {\bibinfo {author} {\bibfnamefont {A.}~\bibnamefont
  {Nishizawa}}, \bibinfo {author} {\bibfnamefont {A.}~\bibnamefont {Taruya}},
  \bibinfo {author} {\bibfnamefont {K.}~\bibnamefont {Hayama}}, \bibinfo
  {author} {\bibfnamefont {S.}~\bibnamefont {Kawamura}}, \ and\ \bibinfo
  {author} {\bibfnamefont {M.-a.}\ \bibnamefont {Sakagami}},\ }\href@noop {}
  {\bibfield  {journal} {\bibinfo  {journal} {Physical Review D}\ }\textbf
  {\bibinfo {volume} {79}},\ \bibinfo {pages} {082002} (\bibinfo {year}
  {2009})}\BibitemShut {NoStop}%
\bibitem [{\citenamefont {Callister}\ \emph {et~al.}(2017)\citenamefont
  {Callister}, \citenamefont {Biscoveanu}, \citenamefont {Christensen},
  \citenamefont {Isi}, \citenamefont {Matas}, \citenamefont {Minazzoli},
  \citenamefont {Regimbau}, \citenamefont {Sakellariadou}, \citenamefont
  {Tasson},\ and\ \citenamefont {Thrane}}]{callister}%
  \BibitemOpen
  \bibfield  {author} {\bibinfo {author} {\bibfnamefont {T.}~\bibnamefont
  {Callister}}, \bibinfo {author} {\bibfnamefont {A.~S.}\ \bibnamefont
  {Biscoveanu}}, \bibinfo {author} {\bibfnamefont {N.}~\bibnamefont
  {Christensen}}, \bibinfo {author} {\bibfnamefont {M.}~\bibnamefont {Isi}},
  \bibinfo {author} {\bibfnamefont {A.}~\bibnamefont {Matas}}, \bibinfo
  {author} {\bibfnamefont {O.}~\bibnamefont {Minazzoli}}, \bibinfo {author}
  {\bibfnamefont {T.}~\bibnamefont {Regimbau}}, \bibinfo {author}
  {\bibfnamefont {M.}~\bibnamefont {Sakellariadou}}, \bibinfo {author}
  {\bibfnamefont {J.}~\bibnamefont {Tasson}}, \ and\ \bibinfo {author}
  {\bibfnamefont {E.}~\bibnamefont {Thrane}},\ }\href@noop {} {\bibfield
  {journal} {\bibinfo  {journal} {Physical Review X}\ }\textbf {\bibinfo
  {volume} {7}},\ \bibinfo {pages} {041058} (\bibinfo {year}
  {2017})}\BibitemShut {NoStop}%
\bibitem [{\citenamefont {Abbott}\ \emph
  {et~al.}(2018{\natexlab{b}})\citenamefont {Abbott} \emph
  {et~al.}}]{asearchfor}%
  \BibitemOpen
  \bibfield  {author} {\bibinfo {author} {\bibfnamefont {B.}~\bibnamefont
  {Abbott}} \emph {et~al.},\ }\href@noop {} {\bibfield  {journal} {\bibinfo
  {journal} {arXiv preprint arXiv:1802.10194}\ } (\bibinfo {year}
  {2018}{\natexlab{b}})}\BibitemShut {NoStop}%
\bibitem [{\citenamefont {Allen}\ and\ \citenamefont {Romano}(1999)}]{allen}%
  \BibitemOpen
  \bibfield  {author} {\bibinfo {author} {\bibfnamefont {B.}~\bibnamefont
  {Allen}}\ and\ \bibinfo {author} {\bibfnamefont {J.~D.}\ \bibnamefont
  {Romano}},\ }\href@noop {} {\bibfield  {journal} {\bibinfo  {journal}
  {Physical Review D}\ }\textbf {\bibinfo {volume} {59}},\ \bibinfo {pages}
  {102001} (\bibinfo {year} {1999})}\BibitemShut {NoStop}%
\bibitem [{\citenamefont {Thorne}(1989)}]{300years}%
  \BibitemOpen
  \bibfield  {author} {\bibinfo {author} {\bibfnamefont {K.}~\bibnamefont
  {Thorne}},\ }\href@noop {} {\emph {\bibinfo {title} {Three hundred years of
  gravitation}}}\ (\bibinfo  {publisher} {Cambridge University Press},\
  \bibinfo {year} {1989})\BibitemShut {NoStop}%
\bibitem [{\citenamefont {Isaacson}(1968)}]{isaacson}%
  \BibitemOpen
  \bibfield  {author} {\bibinfo {author} {\bibfnamefont {R.~A.}\ \bibnamefont
  {Isaacson}},\ }\href@noop {} {\bibfield  {journal} {\bibinfo  {journal}
  {Physical Review}\ }\textbf {\bibinfo {volume} {166}},\ \bibinfo {pages}
  {1263} (\bibinfo {year} {1968})}\BibitemShut {NoStop}%
\bibitem [{\citenamefont {Brunetti}\ \emph {et~al.}(1999)\citenamefont
  {Brunetti}, \citenamefont {Coccia}, \citenamefont {Fafone},\ and\
  \citenamefont {Fucito}}]{brunetti}%
  \BibitemOpen
  \bibfield  {author} {\bibinfo {author} {\bibfnamefont {M.}~\bibnamefont
  {Brunetti}}, \bibinfo {author} {\bibfnamefont {E.}~\bibnamefont {Coccia}},
  \bibinfo {author} {\bibfnamefont {V.}~\bibnamefont {Fafone}}, \ and\ \bibinfo
  {author} {\bibfnamefont {F.}~\bibnamefont {Fucito}},\ }\href@noop {}
  {\bibfield  {journal} {\bibinfo  {journal} {Physical Review D}\ }\textbf
  {\bibinfo {volume} {59}},\ \bibinfo {pages} {044027} (\bibinfo {year}
  {1999})}\BibitemShut {NoStop}%
\bibitem [{\citenamefont {O'Connell}\ and\ \citenamefont
  {Salmona}(1967)}]{connell}%
  \BibitemOpen
  \bibfield  {author} {\bibinfo {author} {\bibfnamefont {R.}~\bibnamefont
  {O'Connell}}\ and\ \bibinfo {author} {\bibfnamefont {A.}~\bibnamefont
  {Salmona}},\ }\href@noop {} {\bibfield  {journal} {\bibinfo  {journal}
  {Physical Review}\ }\textbf {\bibinfo {volume} {160}},\ \bibinfo {pages}
  {1108} (\bibinfo {year} {1967})}\BibitemShut {NoStop}%
\bibitem [{\citenamefont {Hawking}(1972)}]{hawking}%
  \BibitemOpen
  \bibfield  {author} {\bibinfo {author} {\bibfnamefont {S.}~\bibnamefont
  {Hawking}},\ }\href@noop {} {\bibfield  {journal} {\bibinfo  {journal}
  {Communications in Mathematical Physics}\ }\textbf {\bibinfo {volume} {25}},\
  \bibinfo {pages} {167} (\bibinfo {year} {1972})}\BibitemShut {NoStop}%
\bibitem [{\citenamefont {Misner}\ \emph {et~al.}(2017)\citenamefont {Misner},
  \citenamefont {Thorne},\ and\ \citenamefont {Wheeler}}]{mtw}%
  \BibitemOpen
  \bibfield  {author} {\bibinfo {author} {\bibfnamefont {C.~W.}\ \bibnamefont
  {Misner}}, \bibinfo {author} {\bibfnamefont {K.~S.}\ \bibnamefont {Thorne}},
  \ and\ \bibinfo {author} {\bibfnamefont {J.~A.}\ \bibnamefont {Wheeler}},\
  }\href@noop {} {\emph {\bibinfo {title} {Gravitation}}}\ (\bibinfo
  {publisher} {Princeton University Press},\ \bibinfo {year}
  {2017})\BibitemShut {NoStop}%
\bibitem [{\citenamefont {Lattimer}\ and\ \citenamefont
  {Prakash}(2001)}]{lattimer}%
  \BibitemOpen
  \bibfield  {author} {\bibinfo {author} {\bibfnamefont {J.}~\bibnamefont
  {Lattimer}}\ and\ \bibinfo {author} {\bibfnamefont {M.}~\bibnamefont
  {Prakash}},\ }\href@noop {} {\bibfield  {journal} {\bibinfo  {journal} {The
  Astrophysical Journal}\ }\textbf {\bibinfo {volume} {550}},\ \bibinfo {pages}
  {426} (\bibinfo {year} {2001})}\BibitemShut {NoStop}%
\bibitem [{\citenamefont {Phinney}(2001)}]{phinney}%
  \BibitemOpen
  \bibfield  {author} {\bibinfo {author} {\bibfnamefont {E.}~\bibnamefont
  {Phinney}},\ }\href@noop {} {\bibfield  {journal} {\bibinfo  {journal} {arXiv
  preprint astro-ph/0108028}\ } (\bibinfo {year} {2001})}\BibitemShut {NoStop}%
\bibitem [{\citenamefont {Vangioni}\ \emph {et~al.}(2015)\citenamefont
  {Vangioni}, \citenamefont {Olive}, \citenamefont {Prestegard}, \citenamefont
  {Silk}, \citenamefont {Petitjean},\ and\ \citenamefont {Mandic}}]{vangioni}%
  \BibitemOpen
  \bibfield  {author} {\bibinfo {author} {\bibfnamefont {E.}~\bibnamefont
  {Vangioni}}, \bibinfo {author} {\bibfnamefont {K.~A.}\ \bibnamefont {Olive}},
  \bibinfo {author} {\bibfnamefont {T.}~\bibnamefont {Prestegard}}, \bibinfo
  {author} {\bibfnamefont {J.}~\bibnamefont {Silk}}, \bibinfo {author}
  {\bibfnamefont {P.}~\bibnamefont {Petitjean}}, \ and\ \bibinfo {author}
  {\bibfnamefont {V.}~\bibnamefont {Mandic}},\ }\href@noop {} {\bibfield
  {journal} {\bibinfo  {journal} {Monthly Notices of the Royal Astronomical
  Society}\ }\textbf {\bibinfo {volume} {447}},\ \bibinfo {pages} {2575}
  (\bibinfo {year} {2015})}\BibitemShut {NoStop}%
\bibitem [{\citenamefont {Kistler}\ \emph {et~al.}(2013)\citenamefont
  {Kistler}, \citenamefont {Yuksel},\ and\ \citenamefont {Hopkins}}]{kistler}%
  \BibitemOpen
  \bibfield  {author} {\bibinfo {author} {\bibfnamefont {M.~D.}\ \bibnamefont
  {Kistler}}, \bibinfo {author} {\bibfnamefont {H.}~\bibnamefont {Yuksel}}, \
  and\ \bibinfo {author} {\bibfnamefont {A.~M.}\ \bibnamefont {Hopkins}},\
  }\href@noop {} {\bibfield  {journal} {\bibinfo  {journal} {arXiv preprint
  arXiv:1305.1630}\ } (\bibinfo {year} {2013})}\BibitemShut {NoStop}%
\bibitem [{\citenamefont {Zhu}\ \emph {et~al.}(2013)\citenamefont {Zhu},
  \citenamefont {Howell}, \citenamefont {Blair},\ and\ \citenamefont
  {Zhu}}]{zhu}%
  \BibitemOpen
  \bibfield  {author} {\bibinfo {author} {\bibfnamefont {X.-J.}\ \bibnamefont
  {Zhu}}, \bibinfo {author} {\bibfnamefont {E.~J.}\ \bibnamefont {Howell}},
  \bibinfo {author} {\bibfnamefont {D.~G.}\ \bibnamefont {Blair}}, \ and\
  \bibinfo {author} {\bibfnamefont {Z.-H.}\ \bibnamefont {Zhu}},\ }\href@noop
  {} {\bibfield  {journal} {\bibinfo  {journal} {Monthly Notices of the Royal
  astronomical Society}\ }\textbf {\bibinfo {volume} {431}},\ \bibinfo {pages}
  {882} (\bibinfo {year} {2013})}\BibitemShut {NoStop}%
\bibitem [{\citenamefont {Steiner}\ \emph {et~al.}(2013)\citenamefont
  {Steiner}, \citenamefont {Lattimer},\ and\ \citenamefont {Brown}}]{steiner}%
  \BibitemOpen
  \bibfield  {author} {\bibinfo {author} {\bibfnamefont {A.~W.}\ \bibnamefont
  {Steiner}}, \bibinfo {author} {\bibfnamefont {J.~M.}\ \bibnamefont
  {Lattimer}}, \ and\ \bibinfo {author} {\bibfnamefont {E.~F.}\ \bibnamefont
  {Brown}},\ }\href@noop {} {\bibfield  {journal} {\bibinfo  {journal} {The
  Astrophysical Journal Letters}\ }\textbf {\bibinfo {volume} {765}},\ \bibinfo
  {pages} {L5} (\bibinfo {year} {2013})}\BibitemShut {NoStop}%
\bibitem [{\citenamefont {Crocker}\ \emph {et~al.}(2015)\citenamefont
  {Crocker}, \citenamefont {Mandic}, \citenamefont {Regimbau}, \citenamefont
  {Belczynski}, \citenamefont {Gladysz}, \citenamefont {Olive}, \citenamefont
  {Prestegard},\ and\ \citenamefont {Vangioni}}]{crocker}%
  \BibitemOpen
  \bibfield  {author} {\bibinfo {author} {\bibfnamefont {K.}~\bibnamefont
  {Crocker}}, \bibinfo {author} {\bibfnamefont {V.}~\bibnamefont {Mandic}},
  \bibinfo {author} {\bibfnamefont {T.}~\bibnamefont {Regimbau}}, \bibinfo
  {author} {\bibfnamefont {K.}~\bibnamefont {Belczynski}}, \bibinfo {author}
  {\bibfnamefont {W.}~\bibnamefont {Gladysz}}, \bibinfo {author} {\bibfnamefont
  {K.}~\bibnamefont {Olive}}, \bibinfo {author} {\bibfnamefont
  {T.}~\bibnamefont {Prestegard}}, \ and\ \bibinfo {author} {\bibfnamefont
  {E.}~\bibnamefont {Vangioni}},\ }\href@noop {} {\bibfield  {journal}
  {\bibinfo  {journal} {Physical Review D}\ }\textbf {\bibinfo {volume} {92}},\
  \bibinfo {pages} {063005} (\bibinfo {year} {2015})}\BibitemShut {NoStop}%
\bibitem [{\citenamefont {Buonanno}\ \emph {et~al.}(2005)\citenamefont
  {Buonanno}, \citenamefont {Sigl}, \citenamefont {Raffelt}, \citenamefont
  {Janka},\ and\ \citenamefont {M{\"u}ller}}]{buonanno}%
  \BibitemOpen
  \bibfield  {author} {\bibinfo {author} {\bibfnamefont {A.}~\bibnamefont
  {Buonanno}}, \bibinfo {author} {\bibfnamefont {G.}~\bibnamefont {Sigl}},
  \bibinfo {author} {\bibfnamefont {G.~G.}\ \bibnamefont {Raffelt}}, \bibinfo
  {author} {\bibfnamefont {H.-T.}\ \bibnamefont {Janka}}, \ and\ \bibinfo
  {author} {\bibfnamefont {E.}~\bibnamefont {M{\"u}ller}},\ }\href@noop {}
  {\bibfield  {journal} {\bibinfo  {journal} {Physical Review D}\ }\textbf
  {\bibinfo {volume} {72}},\ \bibinfo {pages} {084001} (\bibinfo {year}
  {2005})}\BibitemShut {NoStop}%
\bibitem [{\citenamefont {Shibata}\ \emph {et~al.}(1994)\citenamefont
  {Shibata}, \citenamefont {Nakao},\ and\ \citenamefont {Nakamura}}]{shibata}%
  \BibitemOpen
  \bibfield  {author} {\bibinfo {author} {\bibfnamefont {M.}~\bibnamefont
  {Shibata}}, \bibinfo {author} {\bibfnamefont {K.}~\bibnamefont {Nakao}}, \
  and\ \bibinfo {author} {\bibfnamefont {T.}~\bibnamefont {Nakamura}},\
  }\href@noop {} {\bibfield  {journal} {\bibinfo  {journal} {Physical Review
  D}\ }\textbf {\bibinfo {volume} {50}},\ \bibinfo {pages} {7304} (\bibinfo
  {year} {1994})}\BibitemShut {NoStop}%
\bibitem [{\citenamefont {Du}\ and\ \citenamefont {Nishizawa}(2016)}]{du}%
  \BibitemOpen
  \bibfield  {author} {\bibinfo {author} {\bibfnamefont {S.~M.}\ \bibnamefont
  {Du}}\ and\ \bibinfo {author} {\bibfnamefont {A.}~\bibnamefont {Nishizawa}},\
  }\href@noop {} {\bibfield  {journal} {\bibinfo  {journal} {Physical Review
  D}\ }\textbf {\bibinfo {volume} {94}},\ \bibinfo {pages} {104063} (\bibinfo
  {year} {2016})}\BibitemShut {NoStop}%
\bibitem [{\citenamefont {Smartt}(2009)}]{smartt}%
  \BibitemOpen
  \bibfield  {author} {\bibinfo {author} {\bibfnamefont {S.~J.}\ \bibnamefont
  {Smartt}},\ }\href@noop {} {\bibfield  {journal} {\bibinfo  {journal} {Annual
  Review of Astronomy and Astrophysics}\ }\textbf {\bibinfo {volume} {47}},\
  \bibinfo {pages} {63} (\bibinfo {year} {2009})}\BibitemShut {NoStop}%
\bibitem [{\citenamefont {Oppenheimer}\ and\ \citenamefont
  {Snyder}(1939)}]{oppenheimer}%
  \BibitemOpen
  \bibfield  {author} {\bibinfo {author} {\bibfnamefont {J.~R.}\ \bibnamefont
  {Oppenheimer}}\ and\ \bibinfo {author} {\bibfnamefont {H.}~\bibnamefont
  {Snyder}},\ }\href@noop {} {\bibfield  {journal} {\bibinfo  {journal}
  {Physical Review}\ }\textbf {\bibinfo {volume} {56}},\ \bibinfo {pages} {455}
  (\bibinfo {year} {1939})}\BibitemShut {NoStop}%
\bibitem [{\citenamefont {Ferrari}\ \emph {et~al.}(1999)\citenamefont
  {Ferrari}, \citenamefont {Matarrese},\ and\ \citenamefont
  {Schneider}}]{ferrari}%
  \BibitemOpen
  \bibfield  {author} {\bibinfo {author} {\bibfnamefont {V.}~\bibnamefont
  {Ferrari}}, \bibinfo {author} {\bibfnamefont {S.}~\bibnamefont {Matarrese}},
  \ and\ \bibinfo {author} {\bibfnamefont {R.}~\bibnamefont {Schneider}},\
  }\href@noop {} {\bibfield  {journal} {\bibinfo  {journal} {Monthly Notices of
  the Royal Astronomical Society}\ }\textbf {\bibinfo {volume} {303}},\
  \bibinfo {pages} {247} (\bibinfo {year} {1999})}\BibitemShut {NoStop}%
\bibitem [{\citenamefont {Zhu}\ \emph {et~al.}(2010)\citenamefont {Zhu},
  \citenamefont {Howell},\ and\ \citenamefont {Blair}}]{zhu2010}%
  \BibitemOpen
  \bibfield  {author} {\bibinfo {author} {\bibfnamefont {X.-J.}\ \bibnamefont
  {Zhu}}, \bibinfo {author} {\bibfnamefont {E.}~\bibnamefont {Howell}}, \ and\
  \bibinfo {author} {\bibfnamefont {D.}~\bibnamefont {Blair}},\ }\href@noop {}
  {\bibfield  {journal} {\bibinfo  {journal} {Monthly Notices of the Royal
  Astronomical Society: Letters}\ }\textbf {\bibinfo {volume} {409}},\ \bibinfo
  {pages} {L132} (\bibinfo {year} {2010})}\BibitemShut {NoStop}%
\bibitem [{\citenamefont {Behroozi}\ \emph {et~al.}(2013)\citenamefont
  {Behroozi}, \citenamefont {Wechsler},\ and\ \citenamefont
  {Conroy}}]{behroozi}%
  \BibitemOpen
  \bibfield  {author} {\bibinfo {author} {\bibfnamefont {P.~S.}\ \bibnamefont
  {Behroozi}}, \bibinfo {author} {\bibfnamefont {R.~H.}\ \bibnamefont
  {Wechsler}}, \ and\ \bibinfo {author} {\bibfnamefont {C.}~\bibnamefont
  {Conroy}},\ }\href@noop {} {\bibfield  {journal} {\bibinfo  {journal} {The
  Astrophysical Journal}\ }\textbf {\bibinfo {volume} {770}},\ \bibinfo {pages}
  {57} (\bibinfo {year} {2013})}\BibitemShut {NoStop}%
\bibitem [{\citenamefont {Crowther}\ \emph {et~al.}(2010)\citenamefont
  {Crowther}, \citenamefont {Schurr}, \citenamefont {Mirschi}, \citenamefont
  {Yusof}, \citenamefont {Parker}, \citenamefont {Goodwin},\ and\ \citenamefont
  {Kassim}}]{crowther}%
  \BibitemOpen
  \bibfield  {author} {\bibinfo {author} {\bibfnamefont {P.}~\bibnamefont
  {Crowther}}, \bibinfo {author} {\bibfnamefont {O.}~\bibnamefont {Schurr}},
  \bibinfo {author} {\bibfnamefont {R.}~\bibnamefont {Mirschi}}, \bibinfo
  {author} {\bibfnamefont {N.}~\bibnamefont {Yusof}}, \bibinfo {author}
  {\bibfnamefont {R.}~\bibnamefont {Parker}}, \bibinfo {author} {\bibfnamefont
  {S.}~\bibnamefont {Goodwin}}, \ and\ \bibinfo {author} {\bibfnamefont
  {H.}~\bibnamefont {Kassim}},\ }\href@noop {} {\bibfield  {journal} {\bibinfo
  {journal} {MNRAS}\ }\textbf {\bibinfo {volume} {408}},\ \bibinfo {pages}
  {731} (\bibinfo {year} {2010})}\BibitemShut {NoStop}%
\bibitem [{\citenamefont {Ajith}\ \emph {et~al.}(2008)\citenamefont {Ajith},
  \citenamefont {Babak}, \citenamefont {Chen}, \citenamefont {Hewitson},
  \citenamefont {Krishnan}, \citenamefont {Sintes}, \citenamefont {Whelan},
  \citenamefont {Br{\"u}gmann}, \citenamefont {Diener}, \citenamefont {Dorband}
  \emph {et~al.}}]{ajith}%
  \BibitemOpen
  \bibfield  {author} {\bibinfo {author} {\bibfnamefont {P.}~\bibnamefont
  {Ajith}}, \bibinfo {author} {\bibfnamefont {S.}~\bibnamefont {Babak}},
  \bibinfo {author} {\bibfnamefont {Y.}~\bibnamefont {Chen}}, \bibinfo {author}
  {\bibfnamefont {M.}~\bibnamefont {Hewitson}}, \bibinfo {author}
  {\bibfnamefont {B.}~\bibnamefont {Krishnan}}, \bibinfo {author}
  {\bibfnamefont {A.}~\bibnamefont {Sintes}}, \bibinfo {author} {\bibfnamefont
  {J.~T.}\ \bibnamefont {Whelan}}, \bibinfo {author} {\bibfnamefont
  {B.}~\bibnamefont {Br{\"u}gmann}}, \bibinfo {author} {\bibfnamefont
  {P.}~\bibnamefont {Diener}}, \bibinfo {author} {\bibfnamefont
  {N.}~\bibnamefont {Dorband}},  \emph {et~al.},\ }\href@noop {} {\bibfield
  {journal} {\bibinfo  {journal} {Physical Review D}\ }\textbf {\bibinfo
  {volume} {77}},\ \bibinfo {pages} {104017} (\bibinfo {year}
  {2008})}\BibitemShut {NoStop}%
\bibitem [{voy(2015)}]{voyager}%
  \BibitemOpen
  \href@noop {} {\bibfield  {journal} {\bibinfo  {journal} {LIGO Document
  T1400316}\ } (\bibinfo {year} {2015})}\BibitemShut {NoStop}%
\bibitem [{\citenamefont {Sathyaprakash}\ and\ \citenamefont
  {Schutz}(2009)}]{sathyaprakash}%
  \BibitemOpen
  \bibfield  {author} {\bibinfo {author} {\bibfnamefont {B.~S.}\ \bibnamefont
  {Sathyaprakash}}\ and\ \bibinfo {author} {\bibfnamefont {B.~F.}\ \bibnamefont
  {Schutz}},\ }\href@noop {} {\bibfield  {journal} {\bibinfo  {journal} {Living
  Reviews in Relativity}\ }\textbf {\bibinfo {volume} {12}},\ \bibinfo {pages}
  {2} (\bibinfo {year} {2009})}\BibitemShut {NoStop}%
\bibitem [{\citenamefont {Avilez}\ and\ \citenamefont
  {Skordis}(2014)}]{avilez}%
  \BibitemOpen
  \bibfield  {author} {\bibinfo {author} {\bibfnamefont {A.}~\bibnamefont
  {Avilez}}\ and\ \bibinfo {author} {\bibfnamefont {C.}~\bibnamefont
  {Skordis}},\ }\href@noop {} {\bibfield  {journal} {\bibinfo  {journal}
  {Physical review letters}\ }\textbf {\bibinfo {volume} {113}},\ \bibinfo
  {pages} {011101} (\bibinfo {year} {2014})}\BibitemShut {NoStop}%
\bibitem [{\citenamefont {Bertotti}\ \emph {et~al.}(2003)\citenamefont
  {Bertotti}, \citenamefont {Iess},\ and\ \citenamefont {Tortora}}]{bertotti}%
  \BibitemOpen
  \bibfield  {author} {\bibinfo {author} {\bibfnamefont {B.}~\bibnamefont
  {Bertotti}}, \bibinfo {author} {\bibfnamefont {L.}~\bibnamefont {Iess}}, \
  and\ \bibinfo {author} {\bibfnamefont {P.}~\bibnamefont {Tortora}},\
  }\href@noop {} {\bibfield  {journal} {\bibinfo  {journal} {Nature}\ }\textbf
  {\bibinfo {volume} {425}},\ \bibinfo {pages} {374} (\bibinfo {year}
  {2003})}\BibitemShut {NoStop}%
\bibitem [{\citenamefont {Will}(2006)}]{will}%
  \BibitemOpen
  \bibfield  {author} {\bibinfo {author} {\bibfnamefont {C.~M.}\ \bibnamefont
  {Will}},\ }\href@noop {} {\bibfield  {journal} {\bibinfo  {journal} {Living
  reviews in relativity}\ }\textbf {\bibinfo {volume} {9}},\ \bibinfo {pages}
  {3} (\bibinfo {year} {2006})}\BibitemShut {NoStop}%
\bibitem [{\citenamefont {Damour}\ and\ \citenamefont
  {Esposito-Farese}(1993)}]{damour}%
  \BibitemOpen
  \bibfield  {author} {\bibinfo {author} {\bibfnamefont {T.}~\bibnamefont
  {Damour}}\ and\ \bibinfo {author} {\bibfnamefont {G.}~\bibnamefont
  {Esposito-Farese}},\ }\href@noop {} {\bibfield  {journal} {\bibinfo
  {journal} {Physical Review Letters}\ }\textbf {\bibinfo {volume} {70}},\
  \bibinfo {pages} {2220} (\bibinfo {year} {1993})}\BibitemShut {NoStop}%
\bibitem [{\citenamefont {Novak}(1998)}]{novak}%
  \BibitemOpen
  \bibfield  {author} {\bibinfo {author} {\bibfnamefont {J.}~\bibnamefont
  {Novak}},\ }\href@noop {} {\bibfield  {journal} {\bibinfo  {journal}
  {Physical Review D}\ }\textbf {\bibinfo {volume} {58}},\ \bibinfo {pages}
  {064019} (\bibinfo {year} {1998})}\BibitemShut {NoStop}%
\bibitem [{\citenamefont {Harada}(1998)}]{harada}%
  \BibitemOpen
  \bibfield  {author} {\bibinfo {author} {\bibfnamefont {T.}~\bibnamefont
  {Harada}},\ }\href@noop {} {\bibfield  {journal} {\bibinfo  {journal}
  {Physical Review D}\ }\textbf {\bibinfo {volume} {57}},\ \bibinfo {pages}
  {4802} (\bibinfo {year} {1998})}\BibitemShut {NoStop}%
\end{thebibliography}%
\end{document}